%% file: EXO-17-028_temp.tex
\pdfoutput=1

\documentclass[11pt,twoside,a4paper,cmspaper,final,collab]{cms-tdr}
\def\svnVersion{487084P}\def\cmsCernNoTag{CERN-EP-2018-159}\def\cmsCernDate{\today}\def\cmsMessage{Published in the Journal of High Energy Physics as \href{http://dx.doi.org/10.1007/JHEP01(2019)122}{\doi{10.1007/JHEP01(2019)122.}}}

\begin{document}\cmsNoteHeader{EXO-17-028}

\hyphenation{had-ron-i-za-tion}
\hyphenation{cal-or-i-me-ter}
\hyphenation{de-vices}
\RCS$HeadURL: svn+ssh://svn.cern.ch/reps/tdr2/papers/EXO-17-028/trunk/EXO-17-028.tex $
\RCS$Id: EXO-17-028.tex 486659 2019-01-19 22:23:27Z jskim $
\newlength\cmsFigWidth
\newlength\cmsTabSkip\setlength{\cmsTabSkip}{1ex}
\setlength\cmsFigWidth{0.4\textwidth}
\providecommand{\cmsLeft}{left\xspace}
\providecommand{\cmsRight}{right\xspace}
\providecommand{\CL}{CL\xspace}
\providecommand{\NA}{\ensuremath{\text{---}}}

\providecommand{\cmsTable}[1]{\resizebox{\textwidth}{!}{#1}}
\cmsNoteHeader{EXO-17-028}

\newcommand{\VeN}{\ensuremath{V_{\Pe \N}}\xspace}
\newcommand{\VmN}{\ensuremath{V_{\mu \N}}\xspace}
\newcommand{\VlN}{\ensuremath{V_{\ell \N}}\xspace}
\newcommand{\mnu}{\ensuremath{m_\nu}\xspace}
\newcommand{\mN}{\ensuremath{m_\mathrm{N}}\xspace}
\newcommand{\mW}{\ensuremath{m_\mathrm{\PW}}\xspace}
\newcommand{\mZ}{\ensuremath{m_\mathrm{\PZ}}\xspace}
\newcommand{\N}{\ensuremath{\mathrm{N}}\xspace}
\newcommand{\VeNsq}{\ensuremath{\abs{\VeN}^2}\xspace}
\newcommand{\VmNsq}{\ensuremath{\abs{\VmN}^2}\xspace}
\newcommand{\VlNsq}{\ensuremath{\abs{\VlN}^2}\xspace}
\newcommand{\VemNsq}{\ensuremath{\abs{V_{\Pe \N} V^{*}_{\mu \N}}^2 / ( \VeNsq + \VmNsq )}\xspace}

\newcommand{\MT}{\ensuremath{m_\text{T}}\xspace}
\newcommand{\PTJ}{\ensuremath{p_{\mathrm{T}}^{\mathrm{j}}}\xspace}
\newcommand{\WJ}{\ensuremath{\PW}\ensuremath{\text{+jets}}\xspace}
\newcommand{\ZJ}{\ensuremath{\PZ\text{+}\text{jets}}\xspace}
\newcommand{\WZ}{\ensuremath{\PW\PZ}\xspace}
\newcommand{\ZZ}{\ensuremath{\PZ\PZ}\xspace}
\newcommand{\ttZ}{\ensuremath{\ttbar\PZ}\xspace}
\newcommand{\ttW}{\ensuremath{\ttbar\PW}\xspace}
\newcommand{\ttV}{\ensuremath{\ttbar\mathrm{V}}\xspace}
\newcommand{\ttH}{\ensuremath{\ttbar\PH}\xspace}
\newcommand{\relIsoe}{\ensuremath{I^{\Pe}_\text{rel}}\xspace}
\newcommand{\relIsomu}{\ensuremath{I^{\mu}_\text{rel}}\xspace}
\newcommand{\typeI}{Type-I\xspace}
\newcommand{\metSQst}{\ensuremath{(\ptmiss)^2/S_{\mathrm T}}\xspace}
\newcommand{\ptell}{\ensuremath{\pt^{\ell}}\xspace}
\newcommand{\tautwoone}{\ensuremath{\tau_{21}}\xspace}
\newcommand{\ptk}{\ensuremath{p_{\mathrm{T},k}}\xspace}
\newcommand{\pwjet}{\ensuremath{\PW_{\text{jet}}}\xspace}
\newcommand{\nobs}{\ensuremath{\mathrm{N}_{\text{obs}}}\xspace}
\newcommand{\dxy}{\ensuremath{\mathrm{d}_{xy}}\xspace}

\title{Search for heavy Majorana neutrinos in same-sign dilepton channels in proton-proton collisions at $\sqrt{s} = 13\TeV$}

\date{\today}

\abstract{
A search is performed for a heavy Majorana neutrino (\N), produced in leptonic decay of a $\PW$ boson propagator and decaying into a $\PW$ boson and a lepton, with the CMS detector at the LHC.
The signature used in this search consists of two same-sign leptons, in any flavor combination of electrons and muons, and at least one jet.
The data were collected during 2016 in proton-proton collisions at a center-of-mass energy of 13\TeV, corresponding to an integrated luminosity of 35.9\fbinv.
The results are found to be consistent with the expected standard model background.
Upper limits are set in the mass range between 20 and 1600\GeV in the context of a \typeI seesaw mechanism, on \VeNsq, \VmNsq, and \VemNsq, where \VlN is the matrix element describing the mixing of \N with the standard model neutrino of flavor $\ell = \Pe ,\mu$.
For \N masses between 20 and 1600\GeV, the upper limits on \VlNsq range between $2.3 \times 10^{-5}$ and unity.
These are the most restrictive direct limits for heavy Majorana neutrino masses above 430\GeV.
}

\hypersetup{
pdfauthor={CMS Collaboration},
pdftitle={Search for heavy Majorana neutrinos in same-sign dilepton channels in proton-proton collisions at sqrt(s) = 13 TeV},
pdfsubject={CMS},
pdfkeywords={CMS, beyond standard model, Majorana neutrino}
}

\maketitle

\section{Introduction}
The observation of neutrino oscillations~\cite{pdg}, a mixing between several neutrino flavors, established that at least two of the standard model (SM) neutrinos have nonzero masses and that individual lepton number is violated.
The nonzero masses of the neutrinos are arguably the first evidence for physics beyond the SM.
Upper limits on the neutrino masses have been established from cosmological observations~\cite{pdg}, as well as direct measurements, including those of tritium decays~\cite{PhysRev.97.1234,1966JETP...22..521L}.
The extremely small values of these masses are difficult to explain in models that assume neutrinos to be Dirac particles~\cite{Ma:1998dn,Cai:2017mow}.

The leading theoretical candidate to explain neutrino masses is the so-called ``seesaw'' mechanism~\cite{seesaw1, seesaw2, seesaw3, seesaw4, PhysRevD.22.2227,PhysRevD.24.1232,PhysRevD.25.774,Magg198061,PhysRevD.23.165, seesaw5,PhysRevLett.56.561,PhysRevD.34.1642,1987PhLB..187..303B,Lindner:1996tf},
in which a new heavy Majorana neutrino \N is postulated.
In the seesaw mechanism, the observed small neutrino masses, \mnu, result from the large mass of \N, with $\mnu \sim y^2_\nu v^2/ \mN$.
Here $y_\nu$ is a Yukawa coupling, $v$ is the Higgs vacuum expectation value in the SM, and \mN is the mass of the heavy-neutrino state.
One model that incorporates the seesaw mechanism, and can be probed at the LHC, is the neutrino minimal standard model ($\nu$MSM)~\cite{Appelquist:2002me,Appelquist:2003uu,Asaka:2005an,Asaka:2005pn}.
In this model, the existence of new heavy neutrinos could not only explain the very small masses of the SM neutrinos, but also provide solutions to other problems in cosmology, such as the origin of dark matter or the matter-antimatter asymmetry of the early universe~\cite{Asaka:2005an,Asaka:2005pn}.

In this paper, we present the results of a search for a heavy Majorana neutrino in the $\nu$MSM,
which incorporates new heavy-neutrino states without additional vector bosons.
Searches for heavy Majorana neutrinos at hadron colliders have been proposed by many theoretical models~\cite{Maj_hadCol_1, Maj_hadCol_2, Maj_hadCol_3, Maj_hadCol_4, Maj_hadCol_5}.
Numerous experiments have looked for heavy neutrinos in the mass range from several keV to some hundred GeV, with no evidence seen,
and a summary of the limits on \VlNsq versus \mN for these experiments is given in Ref.~\cite{1367-2630-17-7-075019},
where \VlN is a matrix element describing the mixing between the heavy neutrino and the SM neutrino of flavor $\ell = {\Pe},\; \mu$, or $\tau$.
Direct searches for heavy neutrinos have been performed at the CERN LEP collider~\cite{delphi,l3, l3_2001} and, more recently, at the CERN LHC~\cite{Aaij:2014aba,CMS_NR_mu_2012,CMS_NR_emu_2012,ATLAS_NR_2012,Sirunyan:2018mtv}.
These searches use a model-independent phenomenological approach, assuming that \mN and \VlN are free parameters.

The searches performed by the DELPHI~\cite{delphi} and L3~\cite{l3, l3_2001} Collaborations at LEP looked for the $\Pe^{+} \Pe^{-} \to \N \nu_{\ell}$ process, where $\nu_{\ell}$ is any SM neutrino.
For $\ell = \mu , \tau$ the limits on \VlNsq were set for $\mN < 90\GeV$, while for $\ell = \Pe$ the limits extend to $\mN < 200\GeV$.
Several experiments obtained limits for low neutrino masses ($\mN < 5\GeV$), including the LHCb Collaboration~\cite{Aaij:2014aba} at the LHC, which set limits on the mixing of a heavy neutrino with an SM muon neutrino.
The searches by L3, DELPHI, and LHCb include the possibility of a finite heavy-neutrino lifetime, such that \N decays with a vertex displaced from the interaction point.
In the search reported here, however, it is assumed that \N decays close to the point of production, since in the mass range of this search ($\mN > 20\GeV$)
the decay length is expected to be less than $10^{-10}$\unit{m}~\cite{NOTETao2}.

\begin{figure}[htbp]
\centering
  \includegraphics[width=0.9\textwidth]{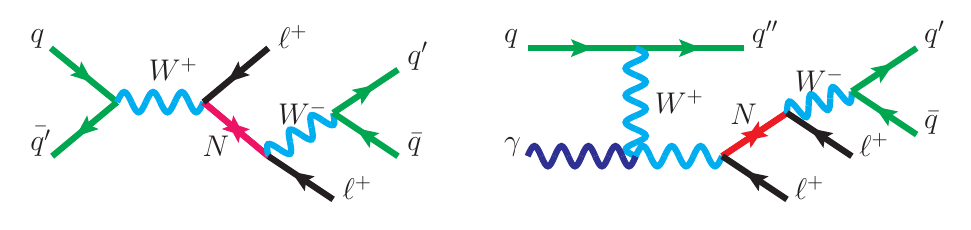}
  \caption{Feynman diagram representing a resonant production of a Majorana neutrino (\N), via the $s$-channel Drell--Yan process (left) and its decay into a lepton and two quarks, resulting in a final state with two same-sign leptons and two quarks from a $\PW$ boson decay.
    Feynman diagram for the photon-initiated process (right).
  }
  \label{fig:schanFD}
\end{figure}

This search probes the decay of a $\PW$ boson, in which an SM neutrino oscillates into a new state \N.
In this analysis, only $\ell = \Pe$ or $\mu$ processes are considered.
In the previous CMS analyses~\cite{CMS_NR_mu_2012,CMS_NR_emu_2012}, only the Drell--Yan (DY) production of \N ($\cPq\cPaq^\prime \to \PW^{*} \to\N \ell^{\pm} \to \ell^{\pm} \ell^{'\pm} \cPq^\prime\cPaq$), shown in Fig.~\ref{fig:schanFD} (left) was considered, while in this study the photon-initiated production of \N ($\cPq \gamma \to \PW \cPq^{\prime\prime} \to \N \ell^{\pm}\cPq^{\prime\prime}\to \ell^{\pm}\ell^{\pm} \cPq^{\prime\prime} \cPq^\prime \cPaq$), as shown in Fig.~\ref{fig:schanFD} (right), is also taken into account.
The diagram in Fig.~\ref{fig:schanFD} (right) shows a possible production of \N via $\PW\gamma$ fusion, which we refer to by the generic term vector boson fusion (VBF).
The inclusion of the VBF channel enhances the sensitivity of this analysis for \N masses above several hundred GeV~\cite{Dev:2013wba}, where the $t$-channel photon-initiated processes become the dominant production mechanism for $\PW^{*} \to \N \ell$~\cite{Alva2015,Dev:2013wba}.

Since \N is a Majorana particle and can decay to a lepton of equal or opposite charge to that of its parent $\PW$ boson, both opposite- and same-sign (SS) lepton pairs can be produced.
This search targets same-sign dilepton (SS2$\ell$) signatures since these final states have very low SM background.
We search for events where the \N decays to a lepton and a $\PW$ boson, and the $\PW$ boson decays hadronically,
as this allows the reconstruction of the mass of the \N without the ambiguity associated with the longitudinal momentum of an SM neutrino.
For the DY channel production, the final state is $\ell^{+} \ell^{'+} \cPq^\prime\cPaq$.
The charge-conjugate decay chain also contributes and results in an $\ell^{-} \ell^{'-} \cPaq^\prime\cPq$ final state.
In the VBF channel, an additional forward jet is produced in the event.

An observation of the $\ell^{\pm} \ell^{'\pm} \cPq^\prime\cPaq (\cPq^{\prime\prime})$ process would constitute direct evidence of lepton number violation.
The study of this process in different dilepton channels improves the likelihood for the discovery of \N, and constrains the mixing matrix elements.
The dielectron ($\Pe \Pe$), dimuon ($\mu\mu$), and electron-muon ($\Pe \mu$) channels are searched for and allow constraints to be set on \VeNsq, \VmNsq, and \VemNsq, respectively~\cite{NOTETao2}.
In the $\Pe \mu$ channel, the leptons from the $\PW$ boson and the \N decay can be either $\Pe$ and $\mu$, or $\mu$ and $\Pe$, respectively, so the branching fraction
for this channel is twice as large as that for the $\Pe\Pe$ or $\mu\mu$ channels.

The most recent CMS search for heavy Majorana neutrinos in events with two leptons and jets was performed for the mass range $\mN = 40$--500\GeV in the $\Pe \Pe$, $\mu\mu$, and $\Pe \mu$ channels at $\sqrt{s} = 8\TeV$~\cite{CMS_NR_mu_2012,CMS_NR_emu_2012}.
A similar search was also performed by the ATLAS Collaboration in the $\Pe \Pe$ and $\mu\mu$ channels~\cite{ATLAS_NR_2012}.
The CMS Collaboration performed a search for heavy Majorana neutrinos in final states with three leptons using the 2016 data set~\cite{Sirunyan:2018mtv}, setting limits on $\VeNsq$ and $\VmNsq$, for the mass range $\mN = 1$--1200\GeV.
In the case of trilepton channels, events that contain both an electron and a muon ($\Pe\Pe\mu,\mu\mu\Pe$)
present an ambiguity about which of the leptons mixes with \N, and it is thus impossible to probe \VemNsq.
This ambiguity is not present in the current analysis with dilepton channels,
allowing limits to be set on \VemNsq.

The CMS analysis at $\sqrt{s} = 8\TeV$ showed that the efficiency for signal events drops for masses above 400\GeV, as a consequence of the Lorentz-boosted topology of the decay products of \N, which causes the signal jets to overlap and be reconstructed as a single jet.
The signal acceptance, which includes the geometrical acceptance and efficiencies of all selection criteria, can be recovered by including events containing a wide jet that is consistent with the process $\PW \to \cPq \cPaq^\prime$, where the decay products of the $\PW$ boson are merged into a single jet~\cite{Das:2017gke}.
It was also observed that the signal acceptance dropped significantly when the mass of \N was below the $\PW$ boson mass (\mW).
For the $\mu\mu$ channel, the signal acceptance was 0.65 (10.9)\% for $\mN = 60\,(125)\GeV$.

For $\mN < \mW$ the final-state leptons and jets are very soft and fail both the trigger- and the analysis-level momentum requirements applied in the 8\TeV analysis.
In the present analysis, cases where one of the signal jets fails the selection criteria are recovered by including events with only one jet.

In this paper, a new search for \N in the $\Pe \Pe$, $\mu\mu$, and $\Pe \mu$ channels is presented using CMS data collected in 2016 at $\sqrt{s} = 13\TeV$.
The enhancement of the signal cross section for $\sqrt{s} = 13\TeV$ compared to $\sqrt{s} = 8\TeV$ is dependent on \mN.
For the cases when \mN is small, \ie, less than 100\GeV, the enhancement of the cross section for signal is similar to that for the background, while at \mN above 1\TeV the increase is nearly an order of magnitude larger than for the background.
The improvement in sensitivity of this analysis, when compared to the 8\TeV analysis, is therefore expected to depend on \mN.
We search for events with two isolated leptons with the same electric charge, with the presence of either a) two or more jets, with no wide jet, b) exactly one jet, with no wide jet, or c) at least one wide jet.
We look for an excess of events above the expected SM background prediction by applying selection criteria to the data to optimize the signal significance for each mass hypothesis.
Heavy Majorana neutrinos with a mass in the range of 20 to 1700\GeV are considered.
There are three potential sources of SS2$\ell$ background:
SM sources in which two prompt SS leptons are produced (a prompt lepton is defined as an electron or muon originating from a $\PW$/$\PZ$/$\gamma^{*}$ boson, \N, or $\tau$ lepton decay), events resulting from misidentified leptons, and opposite-sign dilepton events (\eg, from $\PZ \to \ell^{+} \ell^{-}$, $\PW^{\pm} \PW^{\mp}\to\,\ell^{+}\nu\ell^{-}\overline{\nu}$) in which the sign of one of the leptons is mismeasured.
The last source is negligible for the $\mu \mu$ and $\Pe \mu$ channels.

\section{The CMS detector}
The central feature of the CMS apparatus is a superconducting solenoid of 6\unit{m} internal diameter, providing a magnetic field of 3.8\unit{T}. 
Within the solenoid volume are a silicon pixel and strip tracker, a lead tungstate crystal electromagnetic calorimeter (ECAL), and a brass and scintillator hadron calorimeter (HCAL), each composed of a barrel and two endcap sections.
The ECAL provides a coverage in pseudorapidity $\abs{\eta} < 1.48$ in the barrel region and $1.48 < \abs{\eta} < 3.00$ in the two endcap regions.
Muons are detected in gas-ionization detectors, providing a coverage of $\abs{\eta} < 2.4$, and are embedded in the steel flux-return yoke outside the solenoid.
The first level of the CMS trigger system~\cite{Khachatryan:2016bia}, composed of custom hardware processors, uses information from the calorimeters and muon detectors to select up to 100\unit{kHz} of the most interesting events.
The high-level trigger (HLT) processor farm uses information from all CMS subdetectors to further decrease the event rate to roughly 1\unit{kHz} before data storage.
A more detailed description of the CMS detector can be found in Ref.~\cite{CMS_detector}.

\section{Simulated samples}

Samples of simulated events are used to estimate the background from SM processes containing prompt SS leptons originating from hard-scattering processes and to determine the heavy Majorana neutrino signal acceptance and selection efficiency.
The background from SM sources are produced using the \MGvATNLO~2.2.2 or 2.3.3 Monte Carlo (MC) generator~\cite{Alwall:2014hca} at leading order (LO) or next-to-leading order (NLO) in perturbative quantum chromodynamics (QCD), with the exception of $\Pg\Pg\to\PZ\PZ$ which is simulated at LO with \MCFM 7.0~\cite{Campbell:2010ff},
and the diboson production processes ($\PW\PZ$ and $\PZ\PZ$) that are generated at NLO with the \POWHEG v2~\cite{Nason:2004rx,Frixione:2007vw,Alioli:2010xd,Nason:2013ydw} generator.

The NNPDF3.0~\cite{Ball2015} LO (NLO) parton distribution function (PDF) sets are used for the simulated samples generated at LO (NLO).
For all signal and background samples, showering and hadronization are described using the \PYTHIA 8.212~\cite{Sjostrand:2014zea} generator, with the CUETP8M1~\cite{Khachatryan:2015pea} underlying event tune.
The response of the CMS detector is modeled using \GEANTfour~\cite{Geant4}.
Double counting of the partons generated with \MGvATNLO and \PYTHIA is removed using the MLM~\cite{Alwall:2007fs} and FxFx~\cite{Frederix:2012ps} matching schemes in the LO and NLO samples, respectively.

The \N signals are generated using \MGvATNLO 2.6.0 at NLO precision,
where the decay of \N is simulated with \textsc{MadSpin}~\cite{Artoisenet2013}, following the implementation of Refs.~\cite{PhysRevD.94.053002,Das:2016hof}.
This includes the production of \N via the charged-current DY and VBF processes.
For the charged-current DY production mechanism, we employ the \textsc{NNPDF31\_nnlo\_hessian\_pdfas} PDF set~\cite{Ball2015}, while to include the photon PDF in the VBF ($\PW\gamma$ fusion) mechanism we use the \textsc{LUXqed17\_plus\_PDF4LHC15\_nnlo\_100} PDF set~\cite{PhysRevLett.117.242002}.
The NLO cross section, obtained using the generator at $\sqrt{s} = 13\TeV$, for the DY (VBF) process has a value of 58.3 (0.050)\unit{pb} for $\mN = 40\GeV$, dropping to 0.155 ($9.65 \times 10^{-4}$)\unit{pb} for $\mN = 100\GeV$, and to $9.92 \times 10^{-6}$ ($1.69 \times 10^{-5}$)\unit{pb} for $\mN = 1000\GeV$, assuming $\VlNsq = 0.01$.
The 13 TeV cross section increases by a factor of 1.4 (10) for $\mN = 40\,(1000)\GeV$, when compared to the 8 TeV cross section.
The VBF process becomes the dominant production mode for scenarios where the mass of \N is greater than ${\approx}700\GeV$.
Only the final states with two leptons (electrons or muons) and jets are generated.

Additional $\Pp\Pp$ collisions in the same or adjacent bunch crossings (pileup) are taken into account by superimposing minimum bias interactions simulated with \PYTHIA on the hard-scattering process.
The simulated events are weighted such that the distribution of the number of additional pileup interactions, estimated from the measured instantaneous luminosity for each bunch crossing, matches that in data.
The simulated events are processed with the same reconstruction software as used for the data.

\section{Event reconstruction and object identification}

The reconstructed vertex with the largest value of summed physics-object $\pt^2$ is taken to be the primary $\Pp\Pp$ interaction vertex,
where \pt is the transverse momentum of the physics-objects.
Here the physics objects are the jets, clustered using the jet finding algorithm~\cite{antikt,Cacciari:2011ma}
with the tracks assigned to the vertex as inputs, and the associated missing transverse momentum, \ptmiss, which is defined as the magnitude of the vector sum of the momenta of all reconstructed particles in an event.

The global event reconstruction, based on the particle-flow algorithm~\cite{Sirunyan:2017ulk}, aims to reconstruct and identify each individual particle in an event, with an optimized combination of all subdetector information.
In this process, the identification of the particle type (photon, electron, muon, charged hadron, neutral hadron) plays an important role in the determination of the particle direction and energy.
Photons are identified as ECAL energy clusters not linked to the extrapolation of any charged-particle trajectory to the ECAL.
Electrons are identified as primary charged-particle tracks and potentially several ECAL energy clusters corresponding to this track extrapolation to the ECAL and to possible bremsstrahlung photons emitted along the way through the tracker material.
Muons are identified as tracks in the central tracker consistent with either a track or several hits in the muon system, with no significant associated energy deposits in the calorimeters.
Charged hadrons are identified as charged-particle tracks neither identified as electrons, nor as muons.
Finally, neutral hadrons are identified as HCAL energy clusters not linked to any charged-hadron trajectory, or as ECAL and HCAL energy excesses with respect to the expected charged-hadron energy deposit.

The energy of photons is directly obtained from the ECAL measurement, corrected for zero-suppression effects.
The energy of electrons is determined from a combination of the track momentum at the primary interaction vertex, the corresponding ECAL cluster energy, and the energy sum of all bremsstrahlung photons attached to the track.
The energy of muons is obtained from the corresponding track momentum.
The energy of charged hadrons is determined from a combination of the track momentum and the corresponding ECAL and HCAL energy, corrected for zero-suppression effects and for the response function of the calorimeters to hadronic showers.
Finally, the energy of neutral hadrons is obtained from the corresponding corrected ECAL and HCAL energies.

\subsection{Lepton selection}
Electron candidates are selected in the region $\abs{\eta} < 2.5$, excluding $1.44 < \abs{\eta} < 1.57$.
Their identification is based on a multivariate discriminant built from variables that characterize the shower shape and track quality.
To reject electrons originating from photon conversions in the detector material, electrons must have no measurements missing in the innermost layers of the tracking system and must not be matched to any secondary vertex containing another electron~\cite{Khachatryan:2015hwa}.
To reduce the rate of the electron sign mismeasurement, charges measured from independent techniques are required to be the same, using the ``selective method'' for the charge definition as explained in Ref.~\cite{Khachatryan:2015hwa}, which we refer to as ``tight charge''.
Requiring the electrons to have tight charge reduces the signal efficiency by 1--20\%, depending on \mN, while the background from mismeasured sign is reduced by a factor of 10.
To ensure that electron candidates are consistent with originating from the primary vertex, the transverse (longitudinal) impact parameter of the leptons with respect to this vertex must not exceed 0.1 (0.4)\unit{\mm}.
These electrons must also satisfy $\abs{\dxy}/\sigma(\dxy) < 4$, where $\dxy$ is the transverse impact parameter relative to the primary vertex, estimated from the track fit, and $\sigma(\dxy)$ is its uncertainty.

Muons are selected in the range $\abs{\eta}< 2.4$.
The muon trajectory is required to be compatible with the primary vertex, and to have a sufficient number of hits in the tracker and muon systems.
The transverse (longitudinal) impact parameter of the muons with respect to this vertex must not exceed 0.05 (0.40)\unit{\mm}.
These muons must also satisfy $\abs{\dxy}/\sigma(\dxy) < 3$.

To distinguish between prompt leptons and leptons produced in hadron decays or hadrons misidentified as leptons, a relative isolation variable ($I^{\ell}_{\text{rel}}$) is used.
It is defined for electrons (muons) as the pileup-corrected~\cite{Khachatryan:2015hwa,cmsmuon} scalar $\pt$ sum of the reconstructed charged hadrons originating from the primary vertex, the neutral hadrons, and the photons, within a cone of $\Delta R = \sqrt{\smash[b]{(\Delta\eta)^2 + (\Delta\phi)^2}} = 0.3\,(0.4)$ around the lepton candidate's direction at the vertex, divided by the lepton candidate's \pt.

Electrons that pass all the aforementioned requirements and satisfy $\relIsoe < 0.08$ are referred to as ``tight electrons''.
Electrons that satisfy $\relIsoe < 0.4$, and pass less stringent requirements on the multivariate discriminant and impact parameter are referred to as ``loose electrons''.
Muons that pass all the aforementioned requirements and satisfy $\relIsomu < 0.07$ are referred to as ``tight muons''.
Muons that satisfy $\relIsomu < 0.6$, and pass a less stringent requirement on the impact parameter and track quality requirements are referred to as ``loose muons''.
Electrons within $\Delta R < 0.05$ of a muon are removed, as these particles are likely a photon radiated from the muon.

\subsection{Identification of jets and missing transverse momentum}
For each event, hadronic jets are clustered from the reconstructed particle-flow objects with the infrared and collinear safe anti-\kt jet clustering algorithm~\cite{antikt}, implemented in the \FASTJET package~\cite{Akhmedov:2013hec}.
Two different jet radii, 0.4 and 0.8, are used with this algorithm, producing objects referred to as AK4 and AK8 jets, respectively.
The jet momentum is determined as the vector sum of all particle momenta in the jet, and is found from simulation to be within 5 to 10\% of the true parton momentum over the entire \pt spectrum and detector acceptance.
Additional $\Pp\Pp$ interactions within the same or nearby bunch crossings can contribute additional tracks and calorimetric energy depositions to the jet momentum.
To mitigate this effect, tracks identified to be originating from pileup vertices are discarded and an offset correction is applied to correct for remaining contributions.
Jet energy corrections are derived from simulation to bring the measured response of jets to that of particle level jets on average.
In situ measurements of the momentum balance in dijet, $\text{photon}\text{+jet}$, $\PZ\text{+}\text{jet}$, and multijet events are used to estimate residual differences in jet energy scale in data and simulation, and appropriate corrections are applied~\cite{Khachatryan:2016kdb}.
The jet energy resolution is typically 15\% at 10\GeV, 8\% at 100\GeV, and 4\% at 1\TeV.
Additional selection criteria are applied to remove jets potentially dominated by anomalous contributions from various subdetector components or reconstruction failures.
The AK4 (AK8) jets must have \pt $> 20$ (200)\GeV and $\abs{\eta} < 2.7$ to be considered in the subsequent steps of the analysis.
To suppress jets matched to pileup vertices, AK4 jets must pass a selection based on the jet shape and the number of associated tracks that point to non-primary vertices~\cite{CMS-PAS-JME-16-003}.

The AK8 jets are groomed using a jet pruning algorithm~\cite{PhysRevD.80.051501,PhysRevD.81.094023}: subsequent to the clustering of AK8 jets, their constituents are reclustered with the Cambridge--Aachen algorithm~\cite{Dokshitzer:1997in,Wobisch:1998wt},
where the reclustering sequence is modified to remove soft and wide-angle particles or groups of particles.
This reclustering is controlled by a soft threshold parameter $z_{\text{cut}}$, which is set to 0.1, and an angular separation threshold $\Delta R > m_{\text{jet}}/ p_{\text{T},\text{jet}}$.
The jet pruning algorithm computes the mass of the AK8 jet after removing the soft radiation to provide a better mass resolution for jets, thus improving the signal sensitivity.
The pruned jet mass is defined as the invariant mass associated with the four-momentum of the pruned jet.

In addition to the jet grooming algorithm, the ``$N$-subjettiness'' of jets~\cite{Thaler:2011gf} is used to identify boosted vector bosons that decay hadronically.
This observable measures the distribution of jet constituents relative to candidate subjet axes in order to quantify how well the jet can be divided into $N$ subjets.
Subjet axes are determined by a one-pass optimization procedure that minimizes $N$-subjettiness~\cite{Thaler:2011gf}.
The separation in the phi-azimuth plane between all of the jet constituents and their closest subjet axes are then used to compute the $N$-subjettiness as $\tau_N = 1/d_{0} \Sigma_k \ptk \text{min}(\Delta R_{1,k}, \Delta R_{2,k}, ..., \Delta R_{N,k})$ with the normalization factor $d_{0}= \Sigma_{k} \ptk R_{0}$ where $R_{0}$ is the clustering parameter of the original jet, \ptk is the transverse momentum of the $k$-constituent of the jet and $\Delta R_{N,k} =\sqrt{\smash[b]{(\Delta\eta_{N,k})^2 + (\Delta\phi_{N,k})^2}}$ is its distance to the $N$-th subjet.
In particular, the ratio between $\tau_2$ and $\tau_1$, known as \tautwoone, has excellent capability for separating jets originating from boosted vector bosons from jets originating from quarks and gluons~\cite{Thaler:2011gf}.
To select a high-purity sample of jets originating from a hadronically decaying $\PW$ bosons, the AK8 jets are required to have $\tautwoone < 0.6$ and a pruned jet mass between 40 and 130\GeV.
We refer to these selected jets as $\PW$-tagged jets.
The efficiency of the \tautwoone selection for AK8 jets is measured in a \ttbar-enriched sample in data and simulation.
To correct for observed differences between the estimated and measured efficiencies a scale factor of $1.11\pm0.08$ is applied to the event for each AK8 jet that passes the \tautwoone requirement in the simulation~\cite{CMS-PAS-JME-16-003}.

Identifying jets originating from a bottom quark can help suppress background from \ttbar production.
To identify such jets the combined secondary vertex algorithm~\cite{BTV-16-002} is used.
This algorithm assigns to each jet a likelihood that it contains a bottom hadron, using discriminating variables, such as track impact parameters, the properties of reconstructed decay vertices, and the presence or absence of low-\pt leptons.
The average \cPqb\ tagging efficiency for jets above 20\GeV is 63\%, with an average misidentification probability for light-parton jets of about 1\%.

To avoid double counting due to jets matched geometrically with a lepton, any AK8 jet that is within $\Delta R < 1.0$ of a loose lepton is removed from the event.
Moreover, if an AK4 jet is reconstructed within $\Delta R < 0.4$ of a loose lepton or within $\Delta R < 0.8$ of an AK8 jet, it is not used in the analysis.

The \ptmiss is adjusted to account for the jet energy corrections applied to the event~\cite{Khachatryan:2016kdb}.
The scalar sum of all activity in the event ($S_{\mathrm{T}}$) is used in the selection of our signal region and is defined
as the \pt sum of all AK4 and AK8 jets, leptons, and \ptmiss.
The transverse mass, \MT, a variable used in the suppression of background from leptonic $\PW$ boson decays, is defined as follows:
\begin{linenomath}
\begin{equation}
\MT(\ell,\ptmiss)=  \sqrt { \smash[b]{2 \ptell \ptmiss [1 - \cos(\Delta \phi_{\ell,\ptvecmiss})]}} ,
\end{equation}
\end{linenomath}
where \ptell is the transverse momentum of the lepton and $\Delta \phi_{\ell,\ptvecmiss}$ is the azimuthal angle difference between the lepton momentum and \ptvecmiss vector.

\section{Event selection}

Events used in this search are selected using several triggers, requiring the presence of two charged leptons ($\Pe$ or $\mu$).
All triggers require two loosely isolated leptons,
where the leading- (trailing-)\pt lepton must have
$\pt>23\,(12)\GeV$ for the $\Pe\Pe$,
$\pt>17\,(8)\GeV$ for the $\mu\mu$, and
$\pt>23\,(8)\GeV$ for the $\Pe\mu$ trigger at the HLT stage.
The offline requirements on the leading (trailing) lepton $\pt$ are governed by the trigger thresholds, and are
$\pt > 25\,(15)\GeV$ for the $\Pe \Pe$,
$\pt>20\,(10)\GeV$ for the $\mu\mu$, and $\pt>25\,(10)\GeV$ for the $\Pe \mu$ channels.
The efficiency for signal events to satisfy the trigger in the $\Pe\Pe$, $\mu\mu$, and $\Pe\mu$ channels is above 0.88, 0.94, and 0.88, respectively.

\subsection{Preselection criteria}
At a preselection stage, events are required to contain a pair of SS leptons.
To remove background with soft misidentified leptons, the invariant mass of the dilepton pair is required to be above 10\GeV.
Dielectron events with an invariant mass within 20\GeV of the $\PZ$ boson mass~\cite{pdg} are excluded to reject background from $\PZ$ boson decays in which one electron sign is mismeasured.
In order to suppress background from diboson production, such as \WZ, events with a third lepton identified using a looser set of requirements and with $\pt > 10\GeV$ are removed.
Preselected events are required to have at least one AK4 or one AK8 jet passing the full jet selection.
The same preselection is applied in all three channels ($\Pe\Pe$, $\mu\mu$, $\Pe\mu$).

\subsection{Selection criteria for signal regions}

The kinematic properties of signal events from heavy-neutrino decays depend on its mass.
To distinguish between the two $\PW$ bosons involved in the production and decay sequence, we refer to the $\PW$ boson that produces \N in Fig.~\ref{fig:schanFD} (left) as the $\PW$ boson propagator and the $\PW$ boson that decays to a quark and anti-quark pair as the hadronically decaying $\PW$ boson.
Search regions (SRs) are defined separately for the low-mass and the high-mass hypotheses.
In the low-mass SR ($\mN \leq 80\GeV$), the $\PW$ boson propagator is on-shell and the final-state system of dileptons and two jets should have an invariant mass equal to the $\PW$ boson mass.
In the high-mass SR ($\mN > 80\GeV$), the $\PW$ boson propagator is off-shell but the hadronically decaying $\PW$ boson is on-shell, so the invariant mass of the jets from the hadronically decaying $\PW$ will be consistent with the $\PW$ boson mass.

To maximize the discovery potential over the full mass range,
the low- and high-mass SRs are each further split into regions SR1 and SR2, based on the jet configuration.
The four SRs used in the analysis are defined as:
\begin{itemize}
\item low-mass SR1: number of AK4 jets $\geq$ 2 and number of AK8 jets = 0,
\item high-mass SR1: number of AK4 jets $\geq$ 2 and number of AK8 jets = 0,
\item low-mass SR2: number of AK4 jets = 1 and number of AK8 jets = 0,
\item high-mass SR2: number of AK8 jets $\geq$ 1.
\end{itemize}
Taking the three flavor channels into account, the analysis has 12 separate SRs.

In each SR, the technique of selecting jets associated with the hadronic $\PW$ boson decay is different.
If there are any $\PW$-tagged AK8 jets in the event, the AK8 jet with pruned jet mass closest to \mW is assumed to be from the hadronic $\PW$ boson decay.
For the high-mass SRs, if there are two or more AK4 jets in the event and no AK8 jets, the two AK4 jets with the invariant mass closest to \mW are assigned to the hadronically decaying $\PW$ boson.
In the low-mass SRs, the $\PW$ boson propagator is reconstructed from \N (one lepton + jet(s)) and the additional lepton, and if there are more than two jets, the jets are selected such that the mass is closest to \mW.
If only one jet is reconstructed in the low-mass SR then this is assigned as being from the hadronic $\PW$ boson decay.
The jet(s) assigned to the hadronic $\PW$ boson decay are referred to by the symbol \pwjet to simplify notation in the rest of the paper.

Before optimizing the signal significance for each mass hypothesis we apply a set of loose requirements to select the low- and high-mass SRs.
These requirements are chosen to remove a large fraction of the background while keeping the signal efficiency high.
In the low-mass SRs, the invariant mass of the two leptons and \pwjet is required to be less than 300\GeV.
To remove background from leptonic $\PW$ boson decays, events must have \ptmiss less than 80\GeV.
To remove background from top quark decays, events are vetoed if they contain a \cPqb-tagged AK4 jet.
In the high-mass SRs, the following selections are used.
For SR1 the events are required to have $30 < m(\pwjet) < 150\GeV$ for the invariant mass of the \pwjet and $\PTJ > 25\GeV$, where $\PTJ$ is the \pt of the leading jet.
For SR2 the pruned jet mass must satisfy $40 < m(\pwjet) < 130\GeV$.
Since the \ptmiss is correlated with the energy of the final-state objects,
this requirement is not used in high-mass SRs.
Instead, we use \metSQst, which has a stronger discriminating power between high-mass signal and background.
The \metSQst must be less than 15\GeV.
These selections are summarized in Table~\ref{table:low_highmass}.

\begin{table*}[tb]
\topcaption{
Selection requirements, after applying the preselection criteria, for the low- and high-mass signal regions.
A dash indicates that the variable is not used in the selection.
}
\label{table:low_highmass}
\centering
\begin{tabular}{cccccc}
\hline
\rule{0pt}{2.6ex}
\multirow{2}{*}{Region}  & \ptmiss & \metSQst & $m(\ell^{\pm} \ell^{\pm} \pwjet)$  & $m(\pwjet)$ & $\PTJ$ \\
& (\GeVns) & (\GeVns) & (\GeVns) &  (\GeVns) &  (\GeVns)\\
\hline
Low-mass SR1+SR2  &   $<$80 & \NA & $<$300 &  \NA & $>$20

 \\
High-mass SR1 &   \NA & $<$15  &  \NA   & 30--150 & $>$25 \\
High-mass SR2 &  \NA & $<$15  &  \NA   & 40--130 & $>$200 \\
\hline
\end{tabular}
\end{table*}

\subsubsection{Optimization of signal selection}

After applying the selection criteria in Table~\ref{table:low_highmass}, the signal significance is optimized by combining several different variables using a modified Punzi figure of merit~\cite{punzi}.
The Punzi figure of merit is defined as $\epsilon_{\mathrm{S}}/(a/2 + \delta B)$ where $a$ is the number of standard deviations, and is set equal to 2 to be consistent with the previous CMS analysis, $\epsilon_{\mathrm{S}}$ is the signal selection efficiency, and $\delta B$ is the uncertainty in the estimated background.
The signal regions are optimized separately for each mass hypothesis and for each of the three flavor channels.

The variables used to optimize the signal selection, which are all optimized simultaneously, are:
the transverse momentum of the leading lepton $\pt^{\ell_1}$, and of the trailing lepton $\pt^{\ell_2}$;
the invariant mass of the two leptons and the selected jet(s) $m(\ell^{\pm} \ell^{\pm} \pwjet)$;
the angular separation between the $\pwjet$ and the trailing lepton $\Delta R(\ell_{2}, \pwjet)$;
minimum and maximum requirements on the invariant mass of the lepton (leading or trailing) and the selected jet(s) $m(\ell_{i} \pwjet)$, where $i$=1,2;
and the invariant mass of the two leptons $m(\ell^{\pm} \ell^{\pm})$.
We consider the variable $m(\ell_{i} \pwjet)$, as this should peak at \mN for the signal.
Since it is not known which lepton comes from the \N decay, the event is accepted if either $m(\ell_{i} \pwjet)$ satisfies the requirements.
The optimized window requirements for some SRs are enlarged to give complete coverage of the signal parameter space at negligible loss of sensitivity.
The selection requirements and signal acceptances for each mass hypothesis are summarized later in Section~\ref{sec:results}, in Tables~\ref{table:cutop_Low}--\ref{table:cutop_High_em}, for both low- and high-mass SRs.
Here, the lower efficiency at low \mN is due to the selection requirements on the \pt
of the leptons and jets in a signal with very soft jets and leptons.

\section{Background estimate}

The SM background leading to a final state with two SS leptons and jets are divided into the following categories:

\begin{itemize}
\item \textbf{ SM processes with multiple prompt leptons:} these background are mainly from events with two vector bosons ($\PW^{\pm}\PW^{\pm}$, $\PW\PZ$, $\PZ\PZ$).
We also consider as background a $\PW$ or $\PZ$ boson decaying leptonically and accompanied by radiation of an initial- or final-state photon that subsequently undergoes an asymmetric conversion.
These processes produce a final state that can have three or four leptons.
If one or more of the charged leptons fail the reconstruction or selection criteria these processes can appear to have only two SS leptons.

\item \textbf{ Misidentified leptons}: these are processes that contain one or more leptons that are either misidentified hadrons, are from heavy-flavor jets, from light meson decays, or from a photon in a jet.
These leptons are generally less isolated than a prompt lepton from a $\PW/\PZ$ boson decay and tend to have larger impact parameters.
The main processes with a misidentified lepton in the SRs include \WJ events and \ttbar events, but multijet and DY events also contribute.

\item \textbf{ Sign mismeasurement}: if the signs of leptons are mismeasured in events with jets and two opposite-sign leptons (OS2$\ell$), these events could contaminate a search region.
When the sign of a lepton is mismeasured the lepton will on average have a larger impact parameter in comparison to a lepton from a prompt EW boson decay.
Although the rate of mismeasuring the sign of an electron is small, the abundance of OS2$\ell$ events from DY dilepton production means that this background is significant.
It is suppressed by tight requirements on the impact parameter and on the charge of the electron.
The muon sign mismeasurement rate is known to be negligible, based on studies in simulation and with cosmic ray muons~\cite{muon_charge}, and is not considered in this analysis.
\end{itemize}

\subsection{Background from prompt SS leptons}

Background events that contain two prompt SS leptons are referred to as the prompt-lepton background.
These background are estimated using simulation.
To remove any double counting from the misidentified-lepton background estimate based on control samples in data, the leptons have to originate in the decay of either a $\PW/\PZ/\gamma^{*}$ boson, or a $\tau$ lepton.
The largest contribution comes from \WZ, \ZZ, and asymmetric photon conversions, including those in $\PW\gamma$ and $\PZ\gamma$ events.
The background from \WZ and $\PW\gamma^{*}$ production, with $\PW \to \ell\nu$ and $\PZ (\gamma^{*}) \to \ell\ell$, can yield the same signature as \N production: two SS isolated leptons and jets, when one of the opposite-sign same-flavor (OSSF) leptons is not identified and QCD/pileup jets are reconstructed in the event.
This is the largest prompt contribution in both the low- and high-mass SRs.
This background is estimated from simulation, with the simulated yield normalized to the data in a control region (CR) formed by selecting three tight leptons with $\pt > 25, 15, 10\GeV$ and requiring an OSSF lepton pair with invariant mass $m(\ell^{\pm} \ell^{\mp})$ consistent with the $\PZ$ boson mass: $\abs{m(\ell^{\pm}\ell^{\mp}) - \mZ} < 15\GeV$.
In addition, events are required to have $\ptmiss> 50\GeV$ and $\MT(\ell_{\PW},\ptmiss) > 20\GeV$, where the $\ell_{\PW}$ is the lepton not used in the OSSF pair that is consistent with the $\PZ$ boson.
The ratio of the predicted to observed $\PW\PZ$ background yield in this CR is found to be $1.051 \pm 0.065$.
This factor and its associated uncertainty (both statistical and systematic) is used to normalize the corresponding simulated sample.
The systematic uncertainty on this factor is determined by varying, in the simulation, the properties that are listed in Section~\ref{lab:sim_inc}, by $\pm1$ standard deviation from its central value.

Production of $\PZ\PZ$ events with both $\PZ$ bosons decaying leptonically, with two leptons not identified, results in a possible SS2$\ell$ signature.
This process is estimated from simulation, and the simulated yield is normalized using the CR containing four leptons that form two OSSF lepton pairs with invariant masses consistent with that of the $\PZ$ boson.
The ratio of data to simulation from the CR is found to be $0.979 \pm 0.079$, and is used to normalize the simulated $\PZ\PZ$ sample.
A $\PZ$ boson \pt-dependent EW correction to the cross section~\cite{Bierweiler2013,Gieseke:2014gka,PhysRevD.88.113005} is not included in the simulated samples.
It would correct the cross section by at most 25\%, given the range of $\PZ$ boson \pt probed in this analysis.
Since this correction is larger than the uncertainty on the ratio of data to simulation in the CR, we increase the uncertainty on the normalization to 25\%.

External and internal photon conversions can produce an SS2$\ell$ final state when a photon is produced with a $\PW$ or $\PZ$ boson, and this photon undergoes an asymmetric external or internal conversion ($\gamma^{*} \to \ell^{+}\ell^{-}$) in which one of the leptons has very low \pt and fails the lepton selection criteria.
This background mostly contributes to events in the $\Pe \Pe$ and $\Pe \mu$ channels.
It is obtained from simulation and verified in a data CR enriched in both external and internal conversions from the \ZJ process, with $\PZ \to \ell \ell \gamma^{*}$ and $\gamma^{*} \to \ell \ell$, where one of the leptons is outside the detector acceptance.
The CR is defined by $\abs{m(\ell^{\pm} \ell^{\mp}) - \mZ} > 15\GeV$ and $\abs{m(\ell^{\pm}\ell^{\mp}\ell^{\pm}) - \mZ} < 15\GeV$.
The ratio of data to expected background in the CR is $1.093 \pm 0.075$, and this ratio is used to normalize the MC simulation.

Other rare SM processes that can yield two SS leptons include events from EW production of SS $\PW$ pairs, and double parton scattering,
while any SM process that yields three or more prompt leptons produces SS2$\ell$ final states if one or more of the leptons fails to pass the selection.
Processes in the SM that can yield three or more prompt leptons include triboson processes and \ttbar production associated with a boson (\ttW, \ttZ, and \ttH).
Such processes generally have very small production rates (less than 10\% of total background after the preselection)
and in some cases are further suppressed by the veto on \cPqb-tagged jets and requirements on \ptmiss.
They are estimated from simulation and assigned a conservative uncertainty of 50\%, which accounts for the uncertainties due to experimental effects, event simulation, and theoretical calculations of the cross sections.

\subsection{Background from misidentified leptons}

The most important background source for low-mass signals originates from events containing objects misidentified as prompt leptons.
These originate from {\PB} hadron decays, light-quark or gluon jets, and are typically not well isolated.
Examples of these background include: multijet production, in which one or more jets are misidentified as leptons;
$\PW (\to \ell\nu)$+jets events, in which one of the jets is misidentified as a lepton; and \ttbar decays,
in which one of the top quark decays yields a prompt isolated lepton $(\cPqt \to \PW\cPqb \to \ell\nu \cPqb)$ and the other lepton of same sign arises from a bottom quark decay or a jet misidentified as an isolated prompt lepton.
The simulation is not reliable in estimating the misidentified-lepton background for several reasons,
including the lack of statistically large samples (because of the small probability of a jet to be misidentified as a lepton) and inadequate modeling of the parton showering process.
Therefore, these background are estimated using control samples of collision data.

An independent data sample enriched in multijet events (the ``measurement'' sample) is used to calculate the probability misidentifying a jet that passes minimal lepton selection requirements (``loose leptons'') to also pass the more stringent requirements used to define leptons after the full selection (``tight leptons'').
The misidentification probability is applied as an event-by-event weight to the application sample.
The application sample contains events in which one lepton passes the tight selection, while the other lepton fails the tight selection but passes the loose selection ($N_{n\overline{n}}$), as well as events in which both leptons fail the tight selection, but pass the loose criteria ($N_{\overline{n}\,\overline{n}}$).
The total contribution to the signal regions (\ie, the number of events with both leptons passing the tight selection, $N_{nn}$), is then obtained for each mass hypothesis by
weighting events of type $n\overline{n}$ and $\overline{n}\,\overline{n}$ by the appropriate misidentification probability factors and applying the signal selection requirements to the application sample.
To account for the double counting we correct for $\overline{n}\,\overline{n}$ events that can also be $n\overline{n}$.

The measurement sample is selected by requiring a loose lepton and a jet,
resulting in events that are mostly dijet events, with one jet containing a lepton.
Only one lepton is allowed and requirements of $\ptmiss < 80\GeV$, and $\MT(\ell,\ptmiss) < 25\GeV$ are applied.
The loose lepton and jet are required to be separated in azimuth by $\Delta \phi > 2.5$ and the momentum of the jet is required to be greater than the momentum of the lepton.
These requirements suppress contamination from $\PW$ and $\PZ$ boson decays.
Contamination of prompt leptons in the measurement sample from EW processes is estimated and subtracted using simulation.
The normalization of the prompt lepton simulation is validated in a data sample enriched in \WJ events by requiring events with a single lepton, $\ptmiss > 40\GeV$, and $60 < \MT(\ell,\ptmiss)< 100\GeV$.
The minimum uncertainty that covers the discrepancy between the data and simulation in single-lepton \WJ events (across all $\eta$ and $\pt$ bins considered in the analysis) is 30 (13)\% for electrons (muons) and is assigned as the uncertainty in the prompt lepton normalization.
The larger uncertainty for prompt electron events is to allow for the disagreement between data and simulation in single-electron \WJ events for high-\pt electrons.

The method is validated using a sample of simulated \ttbar, \WJ, and DY events.
The misidentification probabilities used in this validation are obtained from simulated events comprised of jets produced via the strong interaction, referred to as QCD multijet events.
The predicted and observed numbers of events in the $\Pe\Pe$, $\mu\mu$, and $\Pe\mu$ channels agree, at preselection, within 10\% for the \WJ and DY samples, and within 25\% for the \ttbar samples.
The latter figure is reduced to 18\% after rejecting events with a \cPqb-tagged jet.

\subsection{Background from opposite-sign leptons}
\label{lab_OS}
To estimate background due to sign mismeasurement, the probability of mismeasuring the lepton sign is studied.
Only mismeasurement of the electron sign is considered, and this background is estimated only in the $\Pe \Pe$ channel.
The probability of mismeasuring the sign of a prompt electron is obtained from simulated $\PZ \to \Pe^{\pm} \Pe^{\mp}$ events and is parametrized as a function of \pt separately for electrons in the barrel and endcap calorimeters.
The average value and statistical uncertainty for the sign mismeasurement probabilities are found to be $(1.65 \pm 0.12) \times 10^{-5}$ in the inner ECAL barrel region ($\abs{\eta} < 0.8$), $(1.07\pm 0.03) \times 10^{-4}$ in the outer ECAL barrel region ($0.8 < \abs{\eta} < 1.5$), and $(0.63 \pm 0.01) \times 10^{-3}$ in the endcap region.
The sign mismeasurement probabilities are then validated with data separately for the barrel and endcap regions.

To estimate the background due to sign mismeasurement in the $\Pe \Pe$ channel, a weight $W_{p}$ is applied to
data events with all the SR selections considered, except that here the leptons are required to be oppositely signed (OS2$\ell$ events).
$W_{p}$ is given by $W_{p} = p_1 /(1-p_1) + p_2 /(1-p_2)$,
where $p_{1(2)}$ is the probability for the leading (trailing) electron sign to be mismeasured and is determined from simulated events.
The \pt of leptons with a mismeasured sign will be misreconstructed.
To correct for the misreconstructed \pt measurement in the OS2$\ell$ events, the lepton \pt is shifted up by $1.5\pm0.5\%$, which is determined from simulation.

To validate the sign mismeasurement probability for the barrel (endcap) region, a control sample of $\PZ \to \Pe^{\pm} \Pe^{\mp}$ events in the data is selected, requiring both electrons to pass through the barrel (endcap) region and demanding the invariant mass of the electron pair to be between 76 and 106\GeV.
The difference between the observed and predicted numbers of $\Pe^{\pm}\Pe^{\pm}$ events is used as a scale factor to account for the modeling in the simulation.
The observed number of events in the data is determined by fitting the $\PZ$ boson mass peak.
The predicted number of events is determined by weighting the OS2$\ell$ events with the value $W_{p}$.
The scale factors and their associated statistical uncertainties in the barrel and endcap regions are found to be $0.80 \pm 0.03$ and $0.87 \pm 0.03$, respectively.

To validate the combined sign mismeasurement probability and scale factors in the data, a control sample of $\PZ \to \Pe^{\pm} \Pe^{\mp}$ events is again selected,
as described above, but here requiring that one electron is found in the endcap and the other, in the barrel region.
The difference in the predicted and observed numbers of $\Pe^{\pm}\Pe^{\pm}$ events in this sample is 12\%.
The same procedure was performed using $\PZ \to \Pe^{\pm} \Pe^{\mp}$ events in the data but requiring no $\eta$ restrictions on the electrons and requiring that the event has only one jet, yielding an agreement within 10\% between the predicted background and the data.

Prompt leptons and background from sign mismeasurement can contaminate the application sample of the misidentified-lepton background, resulting in an overprediction of this background.
This contamination is removed using simulation.
The contamination from the prompt-lepton background is generally less than 1\%. However, for the background from leptons with sign mismeasurement or leptons from photon conversions, the contamination can be as large as 2\% in the signal region and up to 30\% in CR2, that is enriched in background with mismeasured lepton sign.

\subsection{Validation of background estimates}

To test the validity of the background estimation methods, several signal-free data CRs are defined.
The background estimation method is applied in these regions and the results are compared with the observed yields.
These CRs are used to validate the background separately in each of the three flavor channels and are defined as follows:
\begin{itemize}
\item CR1: (SS2$\ell$), at least one \cPqb-tagged AK4 jet,
\item CR2: (SS2$\ell$), $\Delta R(\ell_{1},\ell_{2}) > 2.5$ and no \cPqb-tagged AK4 jet,
\item CR3: (SS2$\ell$), low-mass SR1 and either $\geq$ 1 \cPqb-tagged jet or $\ptmiss > 100\GeV$,
\item CR4: (SS2$\ell$), low-mass SR2 and either $\geq$ 1 \cPqb-tagged jet or $\ptmiss > 100\GeV$,
\item CR5: (SS2$\ell$), high-mass SR1 and either $\geq$ 1 \cPqb-tagged jet or $\metSQst > 20\GeV$,
\item CR6: (SS2$\ell$), high-mass SR2 and either $\geq$ 1 \cPqb-tagged jet or $\metSQst > 20\GeV$.
\end{itemize}
The numbers of predicted and observed background events in each CR are shown in Table~\ref{tab:bkg_val}.
In the control regions CR1 and CR2, the background estimated from data are dominant and validated in events both with and without \cPqb-tagged jets,
while in the remaining CRs all background are validated in regions that are close to the SRs (the misidentified-lepton background accounts for about 90\% of the total background in CR1 and CR2 and about 50\% across the remaining CRs).
The contribution from signal events is found to be negligible in all control regions, with signal accounting for less than 1\% of the yields in most CRs and at most 5\%, when assuming a coupling consistent with the upper limits from previous results.
In all regions the predictions are in agreement with the observations within the statistical and systematic uncertainties described in Section~\ref{sec:syst}, which is dominated by the 30\% uncertainty in the misidentified-lepton background.
Within each region, the observed distributions of all relevant observables also agree with the predictions, within the uncertainties.

\begin{table}[ptb]
\topcaption{Observed event yields and estimated background in the control regions.
            The uncertainties in the background yields are the sums in quadrature of the statistical and systematic components.}
\centering
\begin{tabular}{c c c c}
\hline
Channel & Control region &   Estimated background & Observed \\
\hline
\multirow{6}{*}{$\Pe\Pe$}
 & CR1  & $\phantom{0.0}366 \pm 73\phantom{0.0}$ & \phantom{0}378 \\
 & CR2  & $\phantom{0.0}690 \pm 100\phantom{.0}$ & \phantom{0}671 \\
 & CR3  & $\phantom{0.0}222 \pm 42\phantom{0.0}$ & \phantom{0}242 \\
 & CR4  & $\phantom{00.0}48 \pm 11\phantom{0.0}$ & \phantom{0}\phantom{0}38 \\
 & CR5  & $\phantom{0.0}334 \pm 56\phantom{0.0}$ & \phantom{0}347 \\
 & CR6  & $\phantom{00}25.7 \pm 4.3\phantom{00}$ & \phantom{0}\phantom{0}28 \\[\cmsTabSkip]
\multirow{6}{*}{$\mu\mu$}
 & CR1  & $\phantom{0.0}880 \pm 230\phantom{.0}$ & \phantom{0}925 \\
 & CR2  & $\phantom{0.0}890 \pm 200\phantom{.0}$ & 1013 \\
 & CR3  & $\phantom{0.0}420 \pm 100\phantom{.0}$ & \phantom{0}439 \\
 & CR4  & $\phantom{0.0}156 \pm 42\phantom{0.0}$ & \phantom{0}174 \\
 & CR5  & $\phantom{0.0}560 \pm 120\phantom{.0}$ & \phantom{0}568 \\
 & CR6  & $\phantom{00}35.1 \pm 7.0\phantom{00}$ & \phantom{0}\phantom{0}38 \\[\cmsTabSkip]
\multirow{6}{*}{$\Pe\mu$}
 & CR1  & $\phantom{.0}1010 \pm 240\phantom{.0}$ & 1106 \\
 & CR2  & $\phantom{.0}1350 \pm 230\phantom{.0}$ & 1403 \\
 & CR3  & $\phantom{0.0}650 \pm 140\phantom{.0}$ & \phantom{0}706 \\
 & CR4  & $\phantom{0.0}143 \pm 32\phantom{0.0}$ & \phantom{0}150 \\
 & CR5  & $\phantom{0.0}920 \pm 180\phantom{.0}$ & \phantom{0}988 \\
 & CR6  & $\phantom{00.0}62 \pm 11\phantom{0.0}$ & \phantom{0}\phantom{0}64 \\
\hline
\end{tabular}
\label{tab:bkg_val}
\end{table}

\section{Systematic uncertainties}
\label{sec:syst}

The estimate of background and signal efficiencies is subject to a number of systematic uncertainties.
Table~\ref{tab:sys_rel} shows the contributions from the uncertainty in the signal and background (for two mass hypotheses, $\mN = 50$ and 500\GeV), expressed as a percentage of the total uncertainty.
The relative sizes of these uncertainties for each type of background and signal, in each SR, are listed in Table~\ref{tab:sys}.

\begin{table}[ptb]
\topcaption{Fractional contributions to the total background systematic uncertainties related to the uncertainties in the prompt SS lepton, misidentified-lepton, and
mismeasured-sign background.
The numbers are for the SR1 (SR2) in the case of \mN= 50 and 500\GeV.}
\centering
\begin{tabular}{ccccc}\hline
Channel &   \mN  & Prompt-lepton & Misidentified-lepton & Mismeasured-sign  \\
        &  (\GeVns)  & (\%) & (\%) & (\%) \\
\hline
\multirow{2}{*}{$\Pe\Pe$} & 50 & 53 (49) & 43 (46) & 4.5 (4.9) \\
                          & 500 & 60 (75) & 3.6 (4.6) & 37 (21) \\[\cmsTabSkip]
\multirow{2}{*}{$\mu\mu$} & 50 & 38 (42) & 62 (58) & \NA \\
                          & 500 & 100 (100) & 0.0 (0.0) & \NA \\[\cmsTabSkip]
\multirow{2}{*}{$\Pe\mu$} & 50 & 52 (45) & 48 (55) & \NA \\
                          & 500 & 99 (100) & 1.3 (0.0) & \NA \\
\hline
\end{tabular}
\label{tab:sys_rel}
\end{table}

\begin{table}[ptb]
\topcaption{Summary of the relative systematic uncertainties in heavy Majorana neutrino signal yields and in the background from prompt SS leptons, both estimated from simulation. The relative systematic uncertainties assigned to the misidentified-lepton and mismeasured-sign background estimated from control regions in data and simulation are also shown.
The uncertainties are given for the low- (high-)mass selections. The range given for each systematic uncertainty source covers the variation across the mass range.
Upper limits are presented for the uncertainty related to the PDF choice in the background estimates, however this source of uncertainty is considered to be accounted for via the normalization uncertainty and was not applied explicitly as an uncertainty in the background.}
\centering
\cmsTable{
\begin{tabular}{lcccccc}
\hline
Source / Channel                               & $\Pe \Pe$ signal &  $\Pe \Pe$ bkgd. &     $\mu \mu$ signal & $\mu \mu$ bkgd.  & $\Pe \mu$ signal &  $\Pe \mu$ bkgd. \\
                                                &    (\%)   & (\%)&    (\%)   & (\%)    & (\%) & (\%) \\
\hline
\multicolumn{1}{c}{\textbf{Simulation}:} & & & & & &\\[\cmsTabSkip]
SM cross section				& \NA                                &\phantom{.}12--14 (15--27)\phantom{.} & \NA                                &\phantom{.}13--18 (22--41)\phantom{.} & \NA                                &\phantom{.}12--14 (16--30)\phantom{.} \\
Jet energy scale				&\phantom{.0}2--5 (0--1)\phantom{.0} &\phantom{.00}2--6 (5--6)\phantom{.00} &\phantom{.0}2--8 (0--1)\phantom{.0} &\phantom{.00}3--5 (4--7)\phantom{.00} &\phantom{.0}1--6 (0--1)\phantom{.0} &\phantom{.00}1--4 (3)\phantom{--.000} \\
Jet energy resolution		&\phantom{.0}1--2 (0--0.3)\phantom{} &\phantom{.00}1--2 (2--6)\phantom{.00} &\phantom{.0}1--2 (0--0.3)\phantom{} &\phantom{0}0--0.8 (1--3)\phantom{.00} &\phantom{--0}0.8 (0--0.3)\phantom{} &\phantom{0}0--0.8 (0--3)\phantom{.00} \\
Jet mass scale					&\phantom{}0--0.3 (0--0.1)\phantom{} &\phantom{.00}0--1 (1--3)\phantom{.00} &\phantom{}0--0.2 (0--0.1)\phantom{} &\phantom{0}0--0.3 (0.7)\phantom{--00} &\phantom{}0--0.1 (0--0.1)\phantom{} &\phantom{0}0--0.2 (0--5)\phantom{.00} \\
Jet mass resolution			&\phantom{}0--0.4 (0--0.3)\phantom{} &\phantom{.00}0--1 (0--2)\phantom{.00} &\phantom{}0--0.1 (0--0.2)\phantom{} &\phantom{0}0--0.1 (0--0.5)\phantom{0} &\phantom{}0--0.4 (0--0.3)\phantom{} &\phantom{0}0--0.4 (0--3)\phantom{.00} \\
Subjettiness						&\phantom{.0}0--1 (0--8)\phantom{.0} &\phantom{0}0--1.0 (1--7)\phantom{.00} &\phantom{}0--0.3 (0--8)\phantom{.0} &\phantom{0}0--0.1 (0--8)\phantom{.00} &\phantom{}0--0.2 (0--8)\phantom{.0} &\phantom{0}0--0.4 (0--8)\phantom{.00} \\
Pileup									&\phantom{.0}2--3 (1)\phantom{--.00} &\phantom{--.000}2 (0--2)\phantom{.00} &\phantom{.0}0--1 (0--1)\phantom{.0} &\phantom{.00}0--1 (0--3)\phantom{.00} &\phantom{--0}0.7 (0.8)\phantom{--0} &\phantom{--.000}2 (2--4)\phantom{.00} \\
Unclustered energy			&\phantom{}0--0.7 (0--0.1)\phantom{} &\phantom{--.000}1 (2--5)\phantom{.00} &\phantom{.0}0--1 (0--0.1)\phantom{} &\phantom{.00}0--1 (3--4)\phantom{.00} &\phantom{}0--0.5 (0--0.1)\phantom{} &\phantom{--00}0.9 (1--2)\phantom{.00} \\
Integrated luminosity		&\phantom{--0}2.5 (2.5)\phantom{--0} &\phantom{--00}2.5 (2.5)\phantom{--00} &\phantom{--0}2.5 (2.5)\phantom{--0} &\phantom{--00}2.5 (2.5)\phantom{--00} &\phantom{--0}2.5 (2.5)\phantom{--0} &\phantom{--00}2.5 (2.5)\phantom{--00} \\
Lepton selection				&\phantom{.0}2--4 (4)\phantom{--.00} &\phantom{.00}2--4 (2--6)\phantom{.00} &\phantom{--.00}3 (3--4)\phantom{.0} &\phantom{--.000}3 (3--5)\phantom{.00} &\phantom{--.00}2 (3)\phantom{--.00} &\phantom{--.000}2 (2--6)\phantom{.00} \\
Trigger selection				&\phantom{.0}3--4 (1)\phantom{--.00} &\phantom{--.000}3 (3--5)\phantom{.00} &\phantom{}0--0.9 (0--0.4)\phantom{} &\phantom{.00}0--1 (0--0.8)\phantom{0} &\phantom{--.00}3 (0--0.2)\phantom{} &\phantom{--.000}3 (2)\phantom{--.000} \\
\cPqb\ tagging					&\phantom{}0--0.8 (0--1)\phantom{.0} &\phantom{--00}0.7 (1)\phantom{--.000} &\phantom{}0--0.5 (0--0.6)\phantom{} &\phantom{.00}0--1 (1--3)\phantom{.00} &\phantom{}0--0.7 (0--0.7)\phantom{} &\phantom{.00}0--1 (1--4)\phantom{.00} \\[\cmsTabSkip]
\multicolumn{1}{c}{\textbf{Theory}:} & & & & & & \\[\cmsTabSkip]
PDF variation						&\phantom{}0--0.7 (0--0.2)\phantom{} & $< 15$ ($< 20$)                      &\phantom{}0--0.7 (0--0.1)\phantom{} &  $< 15$ ($< 20$)                     &\phantom{}0--0.7 (0--0.2)\phantom{} & $< 15$ ($< 20$)  \\
Scale variation					&\phantom{.0}1--5 (0--0.1)\phantom{} & \NA                                  &\phantom{.0}1--4 (0--0.3)\phantom{} & \NA                                  &\phantom{.0}1--5 (0--0.2)\phantom{} & \NA  \\[\cmsTabSkip]
\multicolumn{1}{c}{\textbf{Estimated from data}:} & & & & \\[\cmsTabSkip]
Misidentified leptons		& \NA                                &\phantom{--.00}30 (30)\phantom{--.00} & \NA                                &\phantom{--.00}30 (30)\phantom{--.00} & \NA                                &\phantom{--.00}30 (30)\phantom{--.00} \\
Mismeasured sign				& \NA                                &\phantom{.}29--41 (53--88)\phantom{.} & \NA & \NA  & \NA & \NA \\
\hline
\end{tabular}
}
\label{tab:sys}
\end{table}

\subsection{Background uncertainties}

The main sources of systematic uncertainties are associated with the background estimates.
The largest uncertainty is that related to the misidentified-lepton background.
The systematic uncertainty in this background is determined by observing the change in the background estimate with respect to variations in
isolation requirement (and several other selection criteria) for the loose leptons, modifying the \pt requirement for the away-side jet (the jet that is required to be back-to-back with the lepton in the measurement region).
In addition, uncertainties in the jet flavor dependence of the misidentification probability, and in the prompt-lepton contamination in the measurement region are taken into account.
By combining these sources, a systematic uncertainty of 8.9--20\% is assigned.
This uncertainty depends on the lepton flavor and the SR.
The validity of the prediction of the misidentified lepton background was checked by estimating this background using simulated events alone.
The results disagreed with those obtained from the various CRs by up to 30\%,
and this value is assigned as the systematic uncertainty in this background estimate.

The systematic uncertainties in the mismeasured electron sign background are determined by combining weighted average of the uncertainties in barrel/endcap scale factors from background fits, and the uncertainty on the parameterized sign mismeasurement probabilities.
To evaluate the uncertainties in the sign mismeasurement probability scale factors, we vary the range and the number of bins used in the fitting of the data, as well as the requirement on the subleading lepton \pt, and, when combining all these sources, we assign a systematic uncertainty in the scale factors of 9\%.
The uncertainty in the sign mismeasurement probability arising from the choice of parameterization variables was estimated by considering alternative variables such as \metSQst and \ptmiss.
A variation of up to 11\% was observed.
The background estimate method was tested using only simulation, in which OS2$\ell$ events were weighted using the sign mismeasurement probabilities with no scale factors applied.
The predicted and observed number of events in simulation disagree by up to 7\%, and this value is assigned as another source of systematic uncertainty in estimating the sign mismeasurement background.
The three sources discussed above are combined to give a systematic uncertainty of 16\% on this background.
This uncertainty covers the difference between the predicted and observed numbers of events in both data samples enriched in background with mismeasured electrons as discussed in Section~\ref{lab_OS}.

The simulated sample used to measure the sign mismeasurement probabilities has low statistics for events with electron \pt above 100\GeV.
When combined with the uncertainty related to the low statistics of simulated electrons in bins with high electron \pt, for background from mismeasured electron sign, an overall systematic uncertainty of 29--88\% is assigned, depending on electron $\eta$ and \pt.
The large uncertainty in this background applies only to the cases where the SR has two high-\pt electrons.
The effect on the total systematic uncertainty in the background is at most 5\%.

\subsection{Simulation uncertainties}
\label{lab:sim_inc}
The systematic uncertainties in the normalization of the irreducible SM diboson background are taken from the data CR used to normalize the background.
The assigned uncertainties are 6\% for $\PW\PZ$, 25\% for $\PZ\PZ$ and 8\% for $\PZ\gamma$ and $\PW\gamma$ background.
Since other SM processes that can yield two SS leptons, including triboson, \ttV, and $\PW^{\pm}\PW^{\pm}$, have small background yields in the SR, we assign a conservative uncertainty of 50\%, which includes the uncertainties due to experimental effects, event simulation, and theoretical calculations of the cross sections.
The overall systematic uncertainty in the prompt-lepton background, including the contributions discussed below, is 12--18\% for the low-mass selection
and 16--43\% for the high-mass selection, depending on the lepton channel.
To evaluate the uncertainty due to imperfect knowledge of the integrated luminosity~\cite{CMS-PAS-LUM-17-001}, jet energy/mass scale, jet energy/mass resolution~\cite{Khachatryan:2016kdb}, \cPqb\ tagging~\cite{BTV-16-002}, lepton trigger and selection efficiency, as well as the uncertainty in the total inelastic cross section used in the pileup reweighting procedure in simulation, the input value of each parameter is changed by $\pm1$ standard deviation from its central value.
Energy not clustered in the detector affects the overall \ptmiss scale, resulting in an uncertainty in the event yield due to the upper threshold on \ptmiss.

Further uncertainties in the estimation of the yields of the background and signal arise from the unknown higher-order effects in the theoretical calculations of cross sections, 
and from uncertainties in the knowledge of the proton PDFs. 
The theory uncertainties in the renormalization and factorization scales affect the signal cross section and acceptance.
These are evaluated by independently varying the aforementioned scales up and down by a factor of two relative to their nominal values.
The uncertainty associated with the choice of PDFs is estimated following the PDF4LHC recommendations~\cite{Butterworth:2015oua}.
An upper limit on this uncertainty was added to Table~\ref{tab:sys}, although this uncertainty was not applied explicitly in the results but considered to be accounted for via the normalization uncertainty taken from the normalization control regions.

\section{Results and discussion}
\label{sec:results}
\begin{figure}[!hptb]
\centering
    \includegraphics[width=0.40\textwidth]{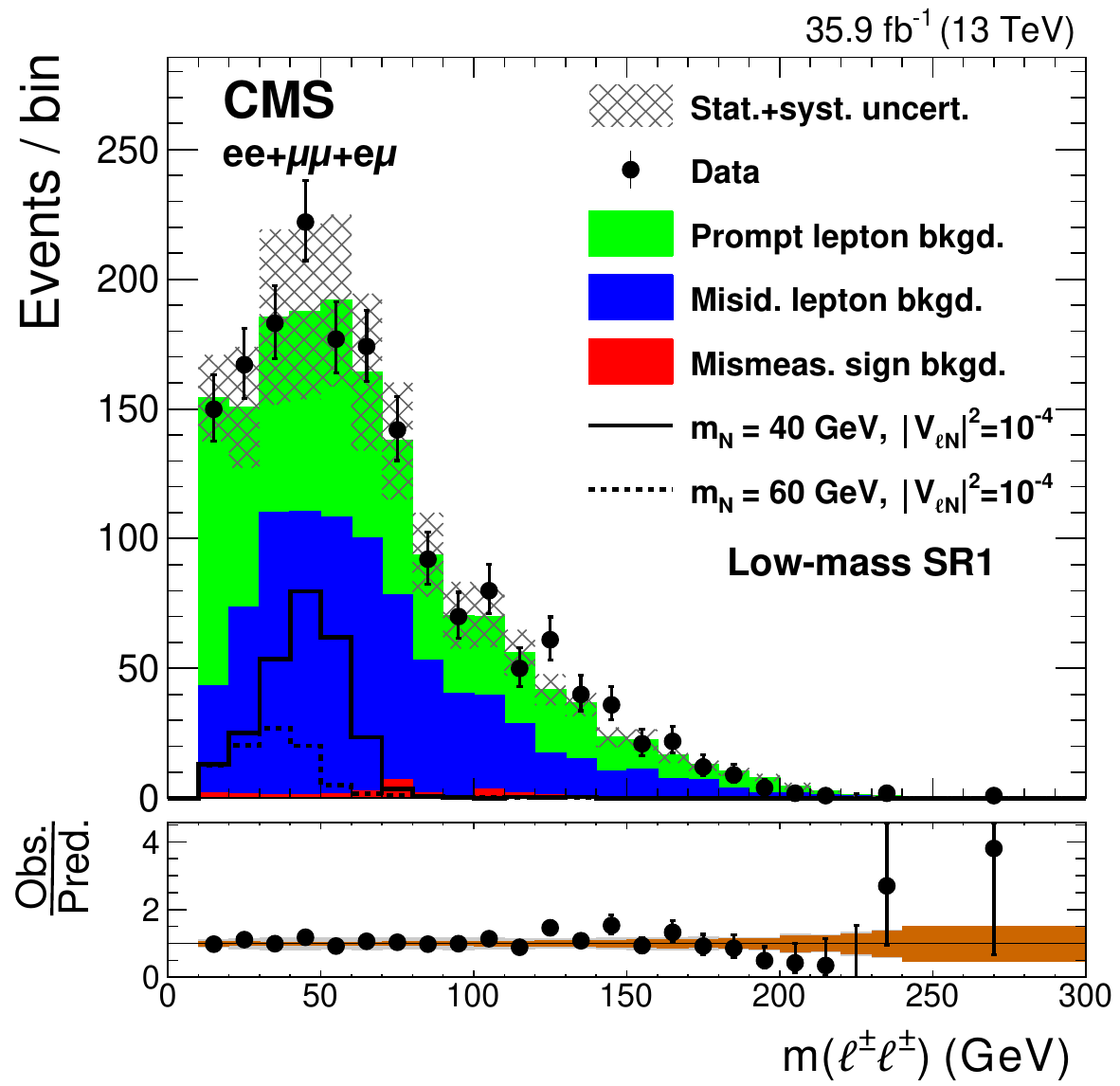}
    \hspace{0.01\textwidth}
    \includegraphics[width=0.40\textwidth]{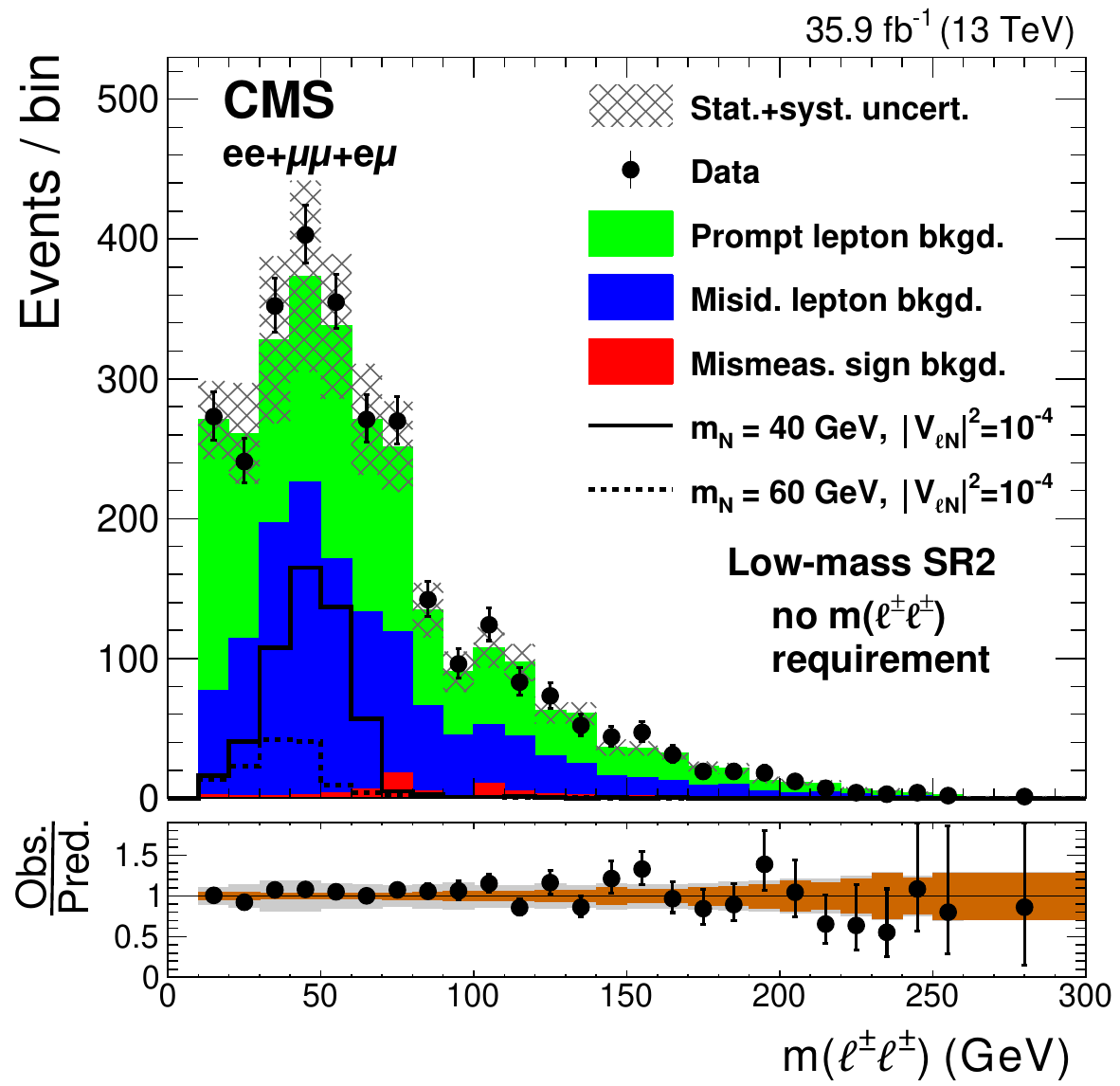}
    \hspace{0.01\textwidth}
    \includegraphics[width=0.40\textwidth]{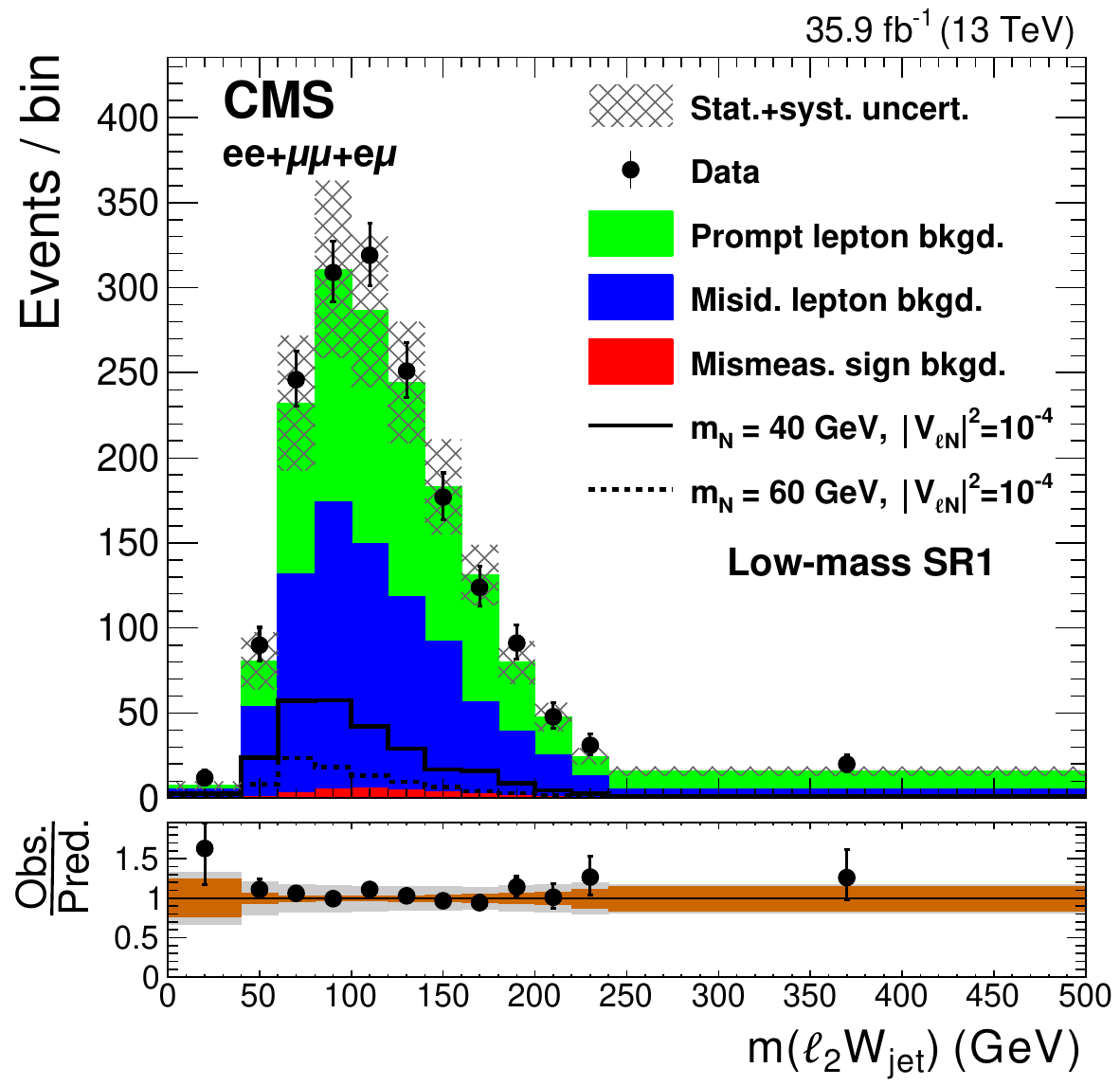}
    \hspace{0.01\textwidth}
    \includegraphics[width=0.40\textwidth]{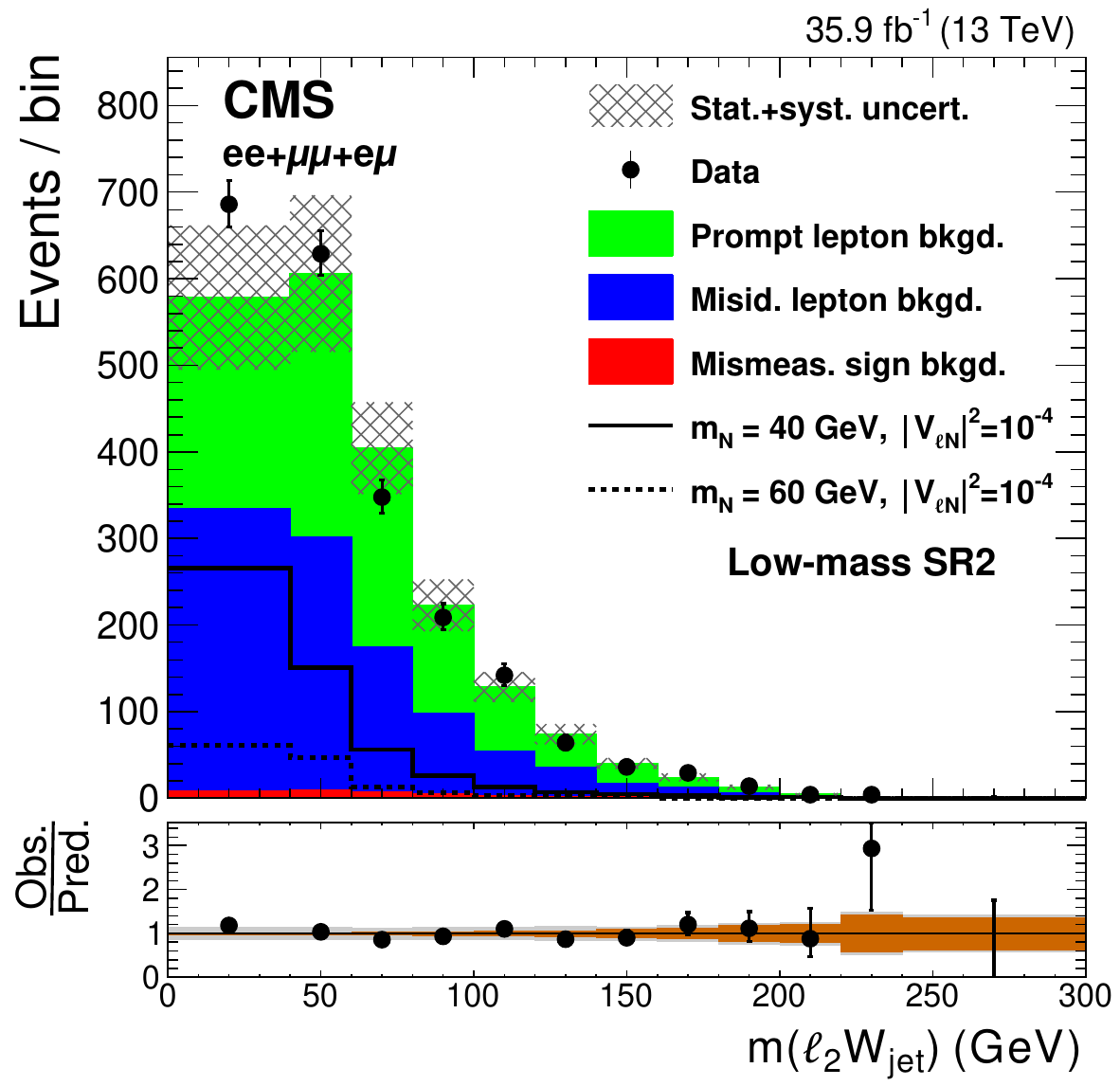}
    \hspace{0.01\textwidth}
    \includegraphics[width=0.40\textwidth]{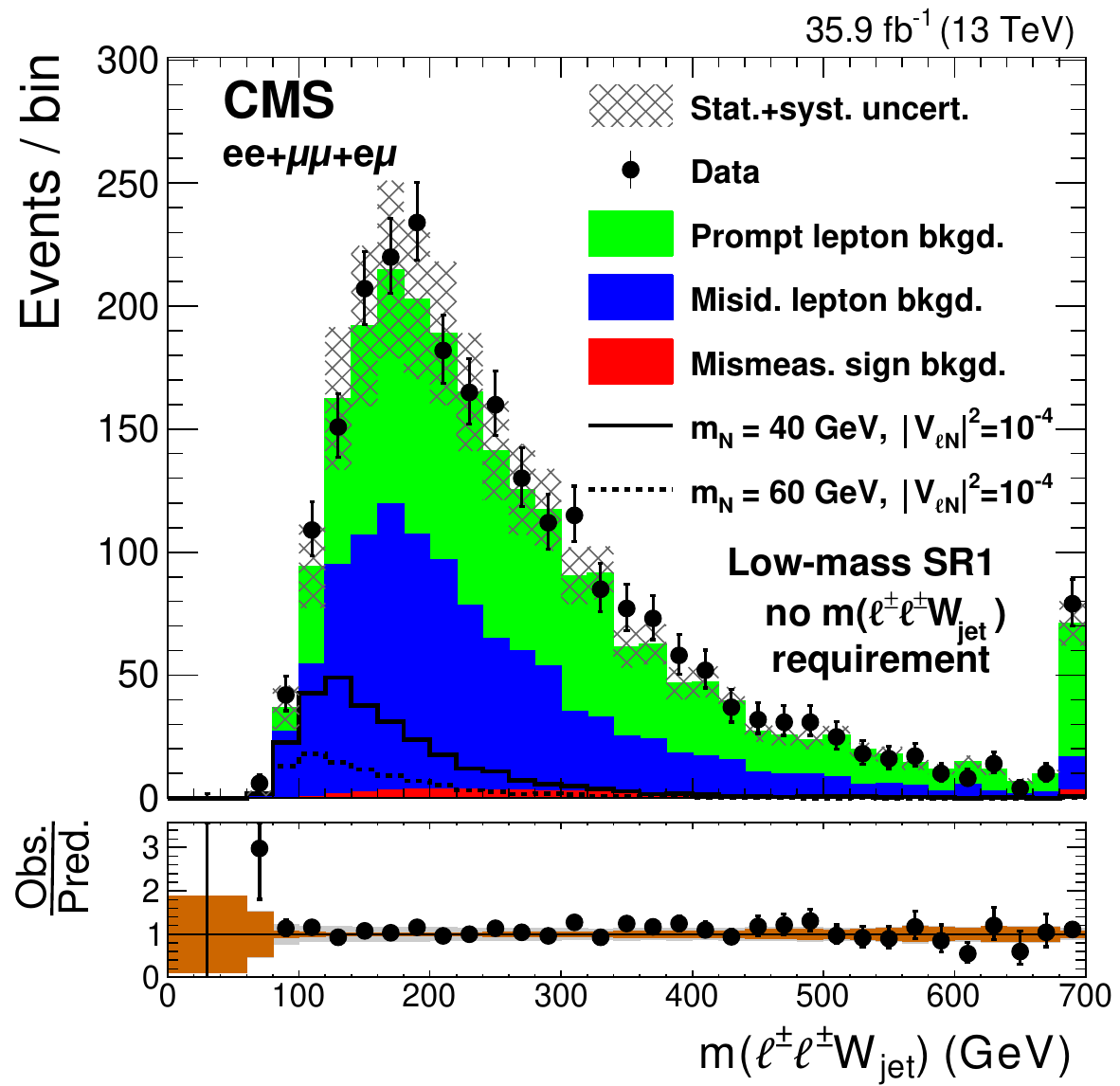}
    \hspace{0.01\textwidth}
    \includegraphics[width=0.40\textwidth]{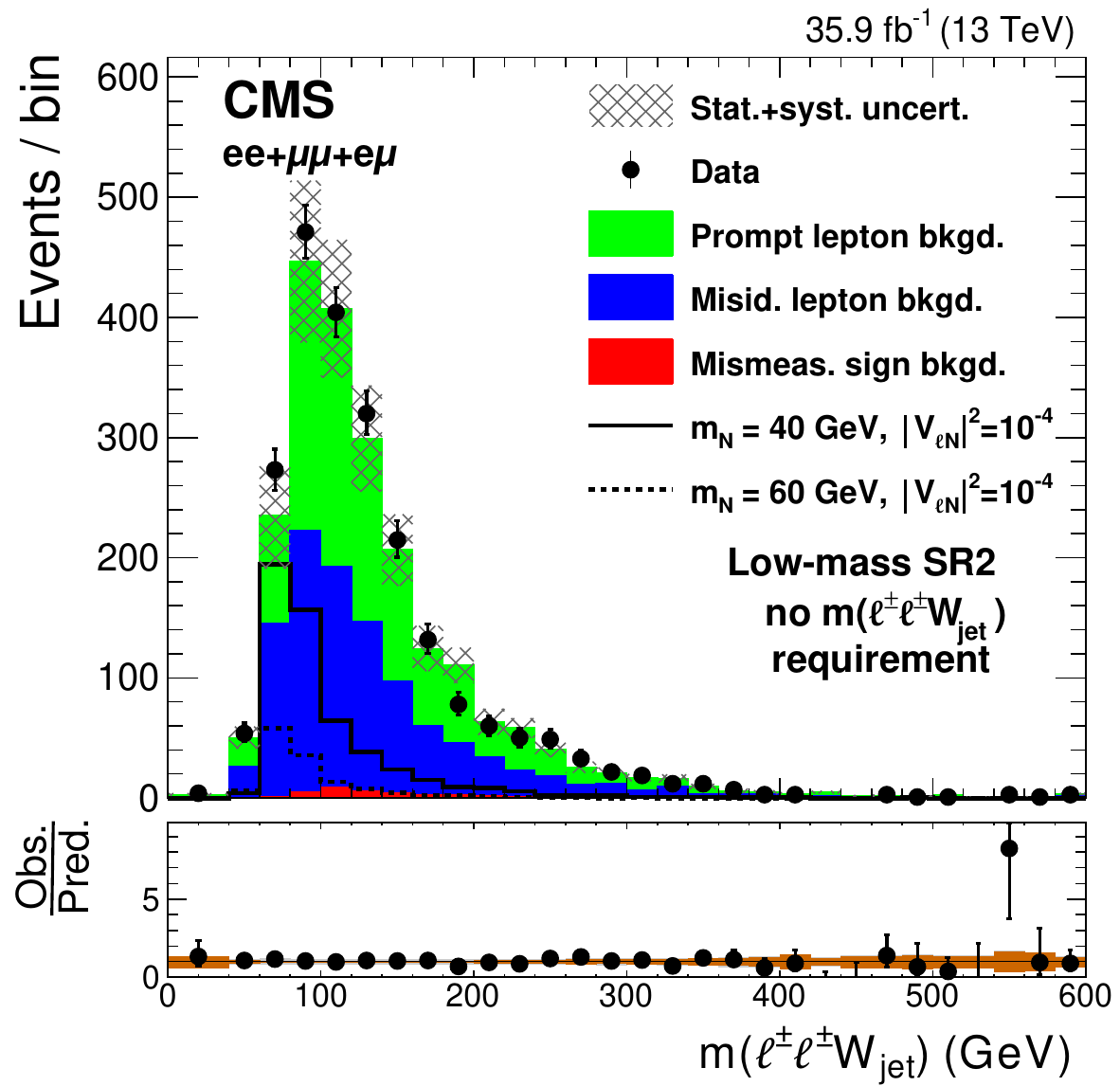}
    \caption{
    Observed distributions of the invariant mass of the two leptons (upper), invariant mass of the subleading lepton and jets (middle), and the invariant mass of the reconstructed $\PW$ propagator (lower), compared to the expected SM background contributions, for the low-mass SR1 (left) and SR2 (right), after combining the events in the $\Pe \Pe$, $\mu \mu$, and $\Pe \mu$ channels.
    The hatched bands represent the sums in quadrature of the statistical and systematic uncertainties.
    The solid and dashed lines show the kinematic distributions of two possible signal hypotheses.
    The lower panels show the ratio between the observed and expected events in each bin, including the uncertainty bands that represent the statistical (brown) and total uncertainties (gray).
    }
    \label{fig:lowmass_SR}
\end{figure}

\begin{figure}[!hptb]
\centering
    \includegraphics[width=0.40\textwidth]{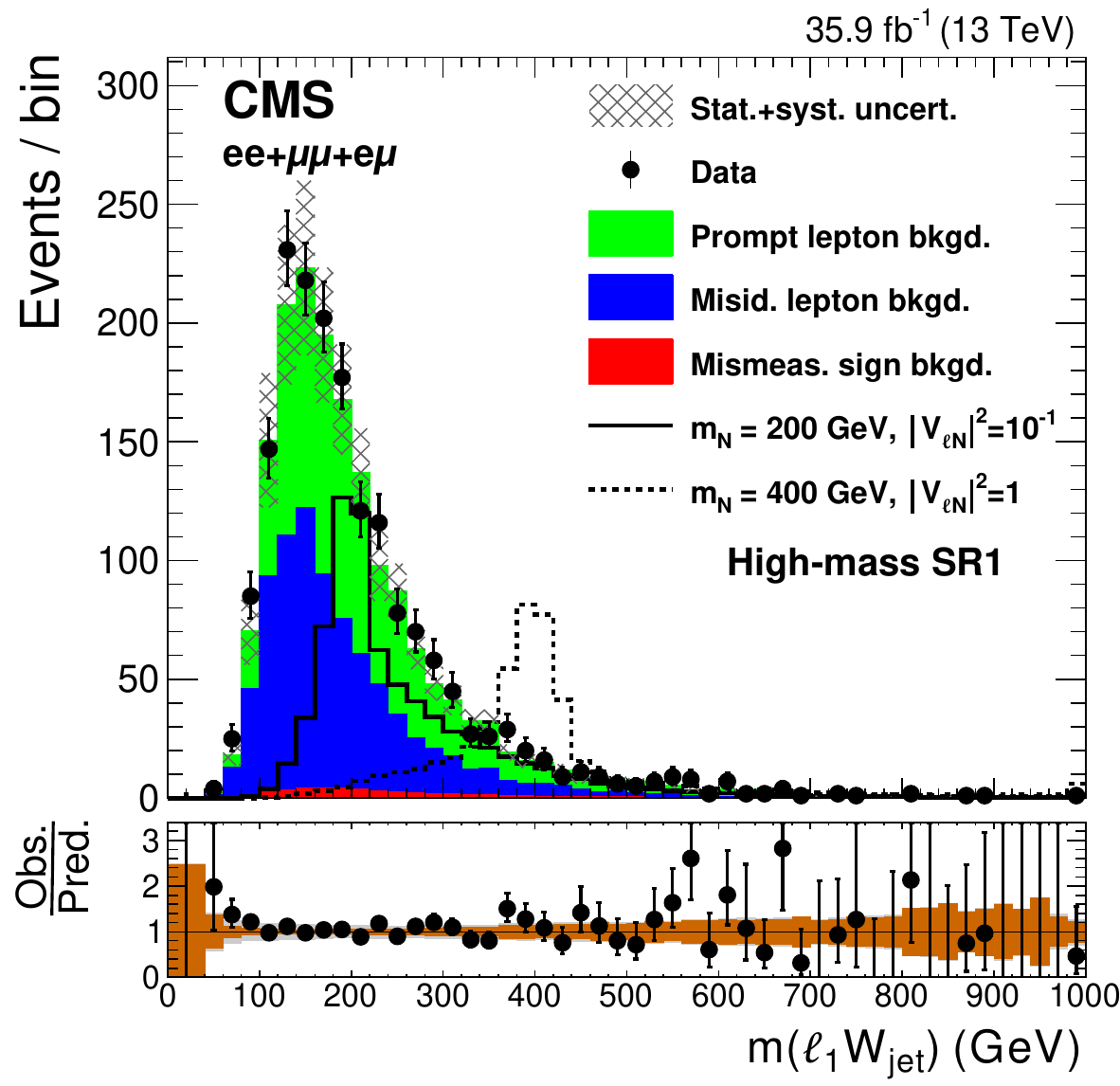}
    \hspace{0.01\textwidth}
    \includegraphics[width=0.40\textwidth]{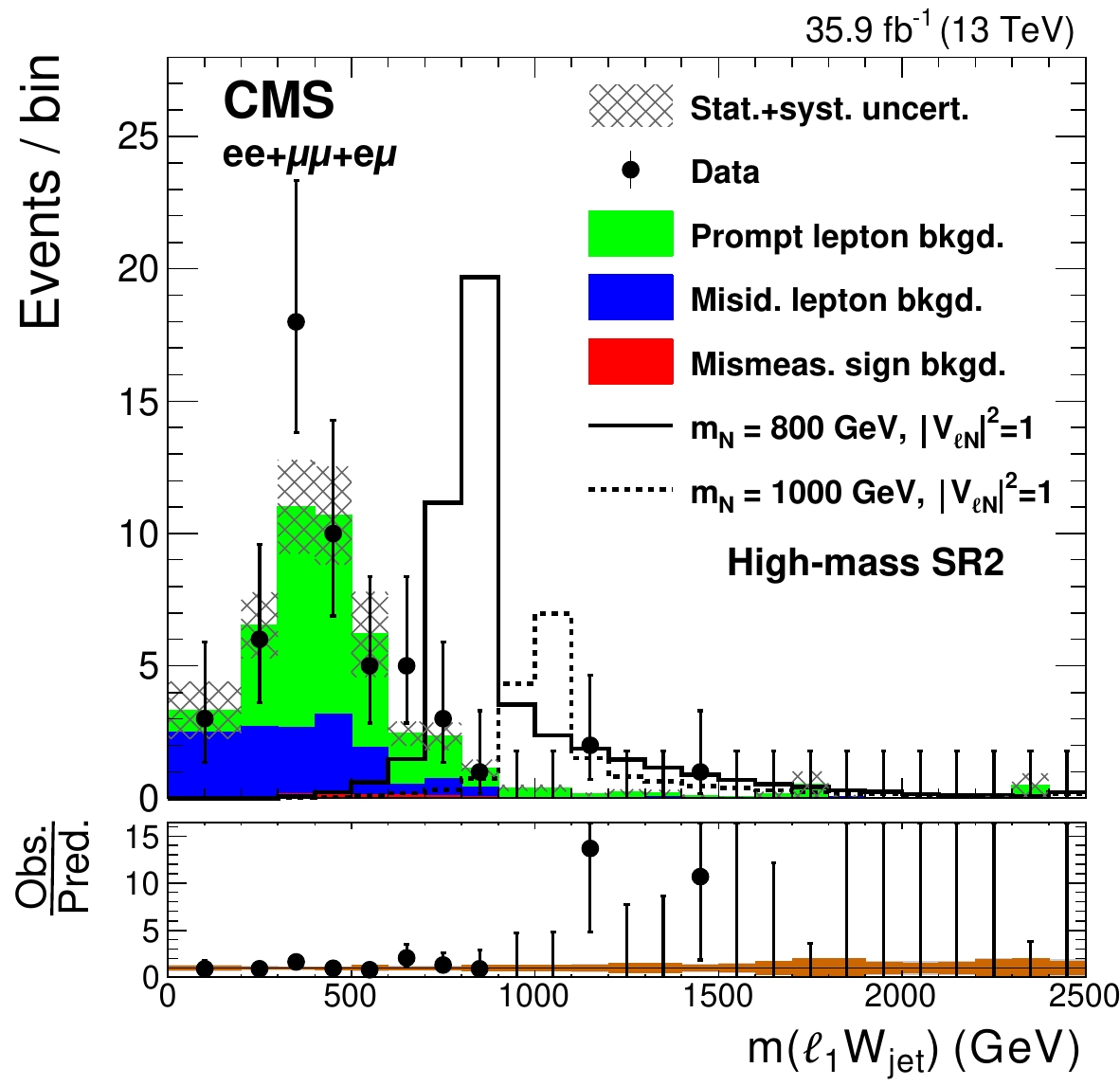}
    \hspace{0.01\textwidth}
    \includegraphics[width=0.40\textwidth]{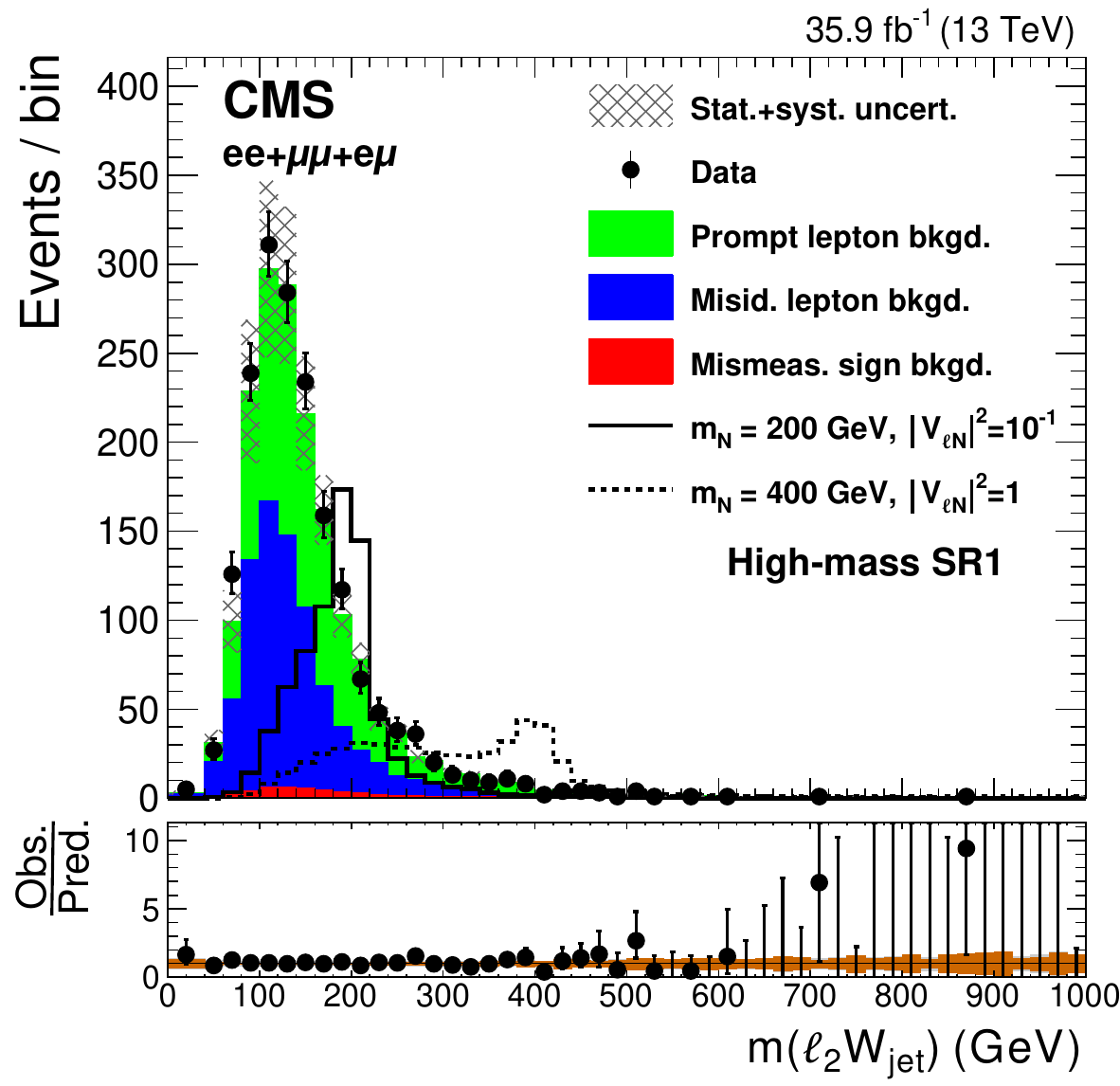}
    \hspace{0.01\textwidth}
    \includegraphics[width=0.40\textwidth]{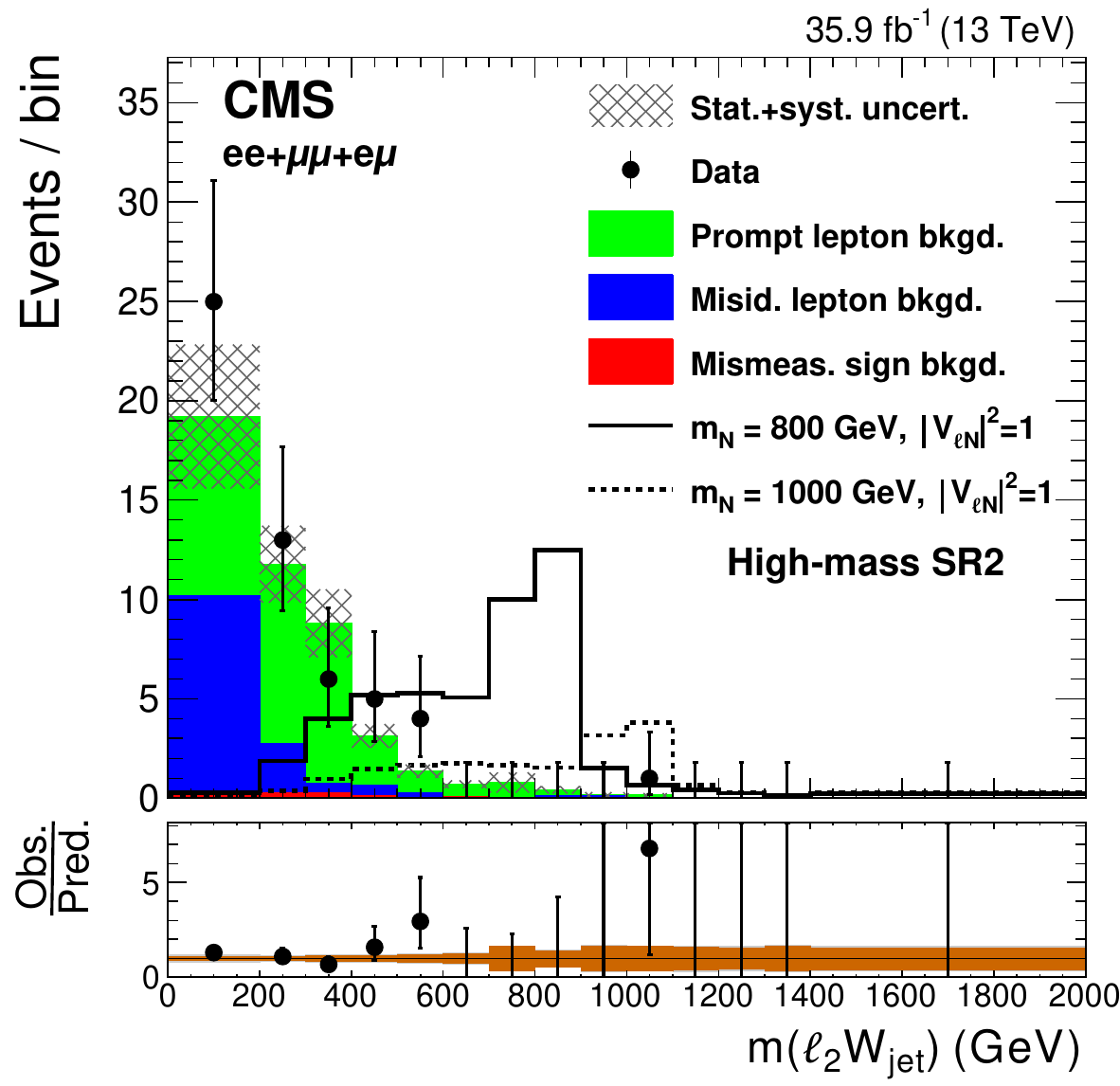}
    \hspace{0.01\textwidth}
    \includegraphics[width=0.40\textwidth]{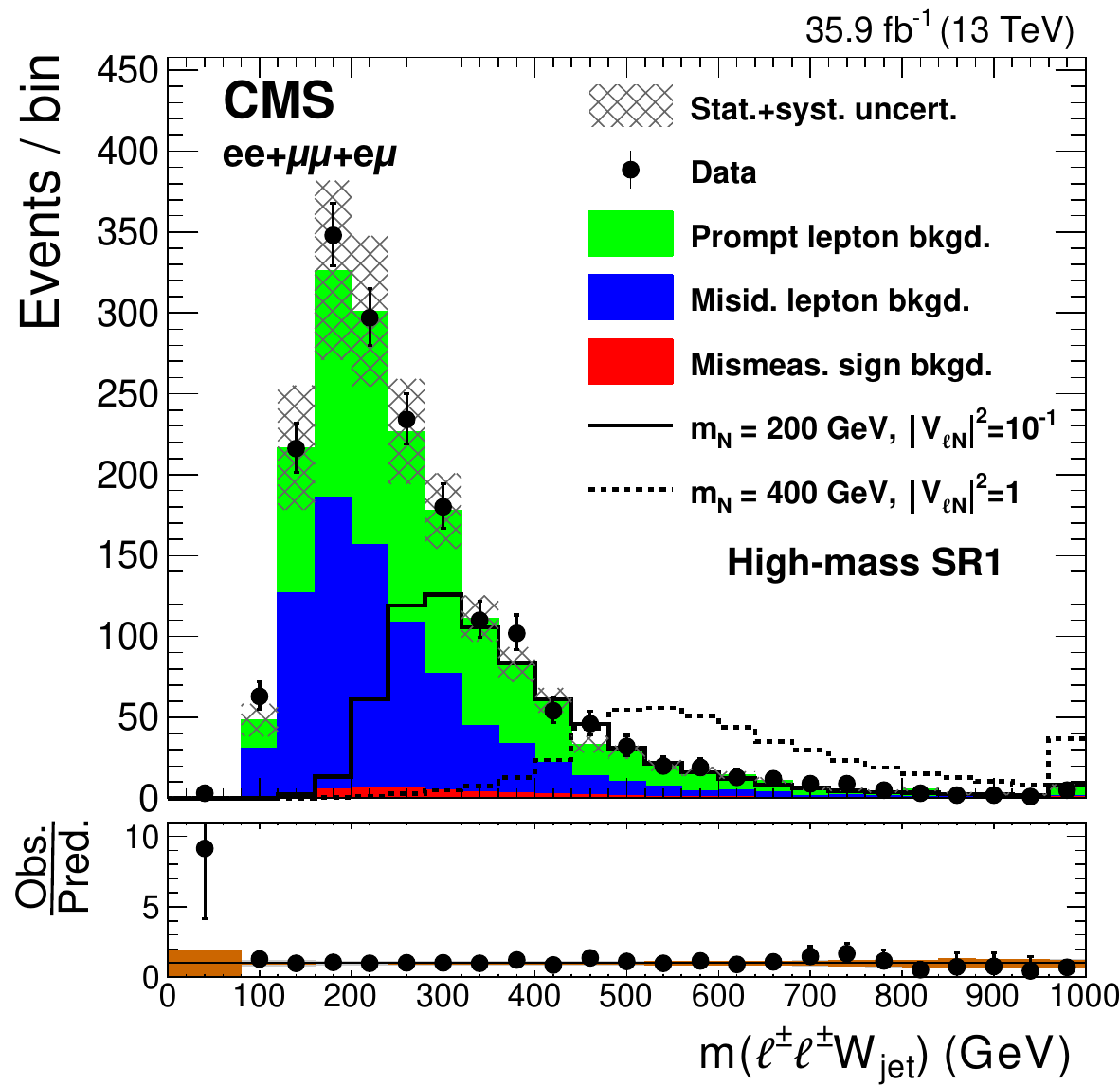}
    \hspace{0.01\textwidth}
    \includegraphics[width=0.40\textwidth]{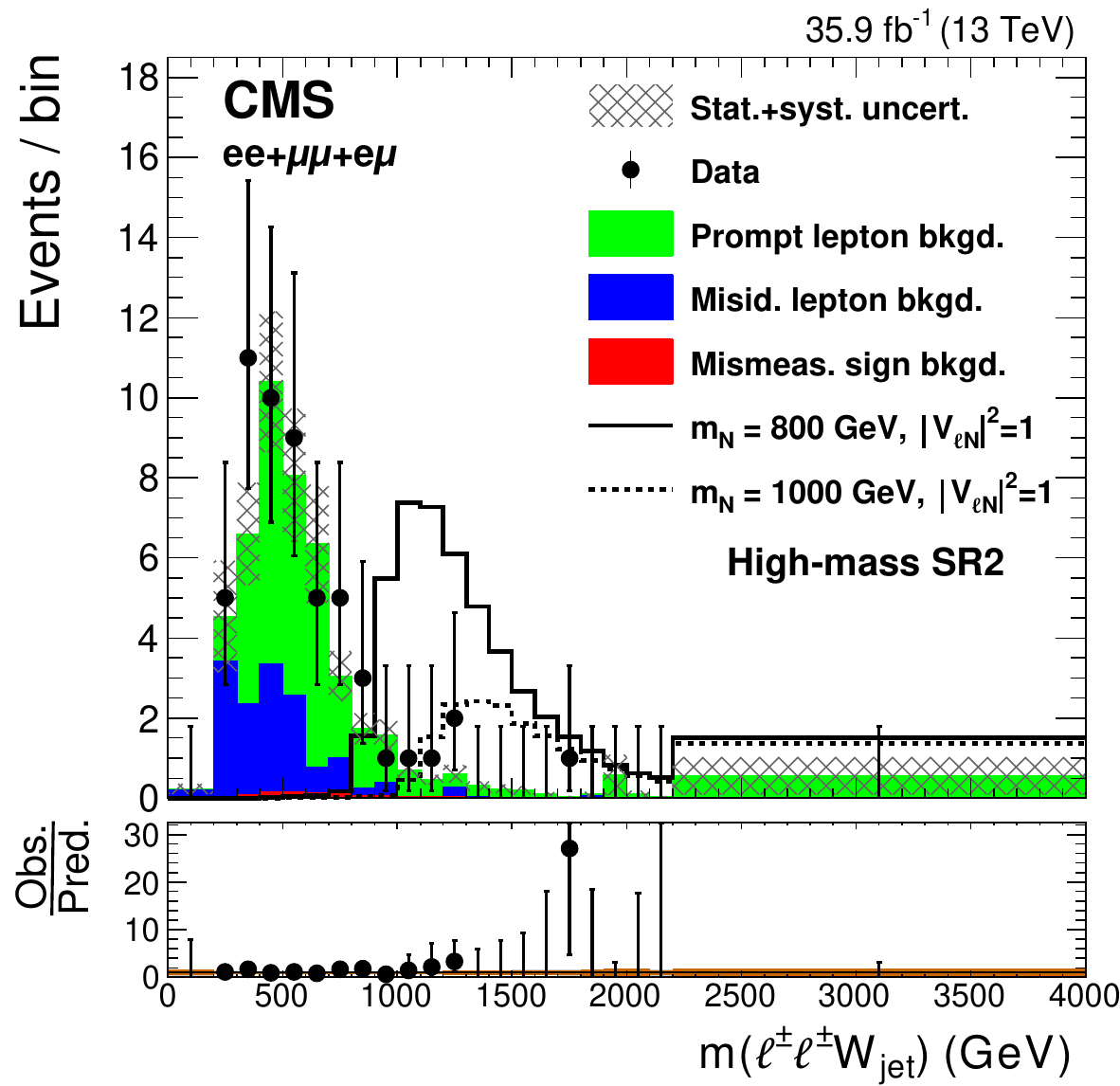}
    \caption{
    Observed distributions of the invariant mass of the leading lepton and jets (upper), invariant mass of the subleading lepton and jets (middle), and the invariant mass of the reconstructed $\PW$ propagator (lower), compared to the expected SM background contributions, for the high-mass SR1 (left) and SR2 (right), after combining the events in the $\Pe \Pe$, $\mu \mu$, and $\Pe \mu$ channels.
    The hatched bands represent the sums in quadrature of the statistical and systematic uncertainties.
    The solid and dashed lines show the kinematic distributions of two possible signal hypotheses.
 The lower panels show the ratio between the observed and expected events in each bin, including the uncertainty bands that represent the statistical (brown) and total uncertainties (gray).
    }
    \label{fig:highmass_SR}
\end{figure}

The data yields and background estimates after the application of the low- and high-mass SR selections are shown in Table~\ref{table:result_SR}.
The predicted background contributed by events with prompt SS leptons, leptons with mismeasured sign,
and misidentified leptons are shown along with the total background estimate and the number of events observed in data.
The uncertainties shown are the statistical and systematic components, respectively.
The data yields are in good agreement with the estimated background.
Kinematic distributions also show good agreement between data and SM expectations.
Figures~\ref{fig:lowmass_SR}--\ref{fig:highmass_SR} show for illustration: the invariant mass of the two leptons (of the leading \pt lepton and the selected jets);
the invariant mass of the trailing \pt lepton and the selected jets;
and the invariant mass of the two leptons and the selected jets for low- (high-)mass SRs.
In Fig.~\ref{fig:lowmass_SR}, the $m(\ell^{\pm} \ell^{\pm} \mathrm{jj})$ signal distribution peaks somewhat below \mW, because of the selection requirements imposed.

\begin{table*}[!tb]
  \topcaption{
Observed event yields and estimated background for the signal region selections.
The background predictions from prompt SS leptons,
misidentified leptons,
leptons with mismeasured sign, and
the total background
are shown together with the number of events observed in data.
The uncertainties shown are the statistical and systematic components, respectively.
A dash indicates that the background is considered negligible.
  }
  \label{table:result_SR}
  \centering
  \cmsTable{
    \begin{tabular}{lccccr}
\hline
SR & Prompt-lepton & Misidentified-lepton & Mismeasured-sign & Total bkgd. & \nobs \\
\hline
\multicolumn{1}{c}{\textbf{\boldmath{$\Pe\Pe$} channel}} \\
Low-mass SR1 &  $206 \pm 10 \pm 21\phantom{0}$  &  $128 \pm 5 \pm 38\phantom{0}$  &  $29.8 \pm 0.2 \pm 12.3$  &  $364 \pm 11 \pm 45\phantom{0}$   & 324 \\
Low-mass SR2 &  $281 \pm 12 \pm 28\phantom{0}$  &  $143 \pm 7 \pm 43\phantom{0}$  &  $36.4 \pm 0.2 \pm 10.7$  &  $461 \pm 14 \pm 53\phantom{0}$   & 460 \\
High-mass SR1 &  $236 \pm 10 \pm 25\phantom{0}$  &  $141 \pm 6 \pm 42\phantom{0}$  &  $45.2 \pm 0.3 \pm 24.0$  &  $422 \pm 12 \pm 55\phantom{0}$   & 382 \\
High-mass SR2 &  $8.0 \pm 1.3 \pm 1.6$  &  $2.0 \pm 0.6 \pm 0.6$  &  $0.91 \pm 0.05 \pm 0.80$ &  $10.9 \pm 1.5 \pm 1.9\phantom{0}$   & 10 \\
\multicolumn{1}{c}{\textbf{\boldmath{$\mu\mu$} channel}} \\
Low-mass SR1 &  $151 \pm 6 \pm 16\phantom{0}$  &  $276 \pm 7 \pm 83\phantom{0}$  & \NA &  $426 \pm 9 \pm 84\phantom{0}$   & 487 \\
Low-mass SR2 &  $209 \pm 8 \pm 19\phantom{0}$  &  $393 \pm 9 \pm 118$  & \NA &  $602 \pm 12 \pm 120$   & 663 \\
High-mass SR1 &  $166 \pm 6 \pm 20\phantom{0}$  &  $244 \pm 6 \pm 73\phantom{0}$  & \NA &  $410 \pm 9 \pm 76\phantom{0}$   & 502 \\
High-mass SR2 &  $7.1 \pm 0.8 \pm 1.9$  &  $4.4 \pm 0.8 \pm 1.3$  & \NA &  $11.5 \pm 1.1 \pm 2.3\phantom{0}$   & 13 \\
\multicolumn{1}{c}{\textbf{\boldmath{$\Pe\mu$} channel}} \\
Low-mass SR1 &  $418 \pm 13 \pm 37\phantom{0}$  &  $432 \pm 10 \pm 130$  & \NA &  $850 \pm 17 \pm 135$   & 907 \\
Low-mass SR2 &  $566 \pm 17 \pm 47\phantom{0}$  &  $464 \pm 12 \pm 139$  & \NA &  $1031 \pm 21 \pm 147\phantom{0}$   & 1042 \\
High-mass SR1 &  $463 \pm 14 \pm 42\phantom{0}$  &  $409 \pm 10 \pm 123$  & \NA &  $871 \pm 17 \pm 129$   & 901 \\
High-mass SR2 &  $16.8 \pm 1.9 \pm 3.6\phantom{0}$  &  $7.4 \pm 1.3 \pm 2.2$  & \NA &  $24.2 \pm 2.3 \pm 4.2\phantom{0}$   & 31 \\
\hline
    \end{tabular}
  }
\end{table*}

The expected signal depends on both \mN and the mixing matrix elements \VeNsq, \VmNsq, or \VemNsq, and the values are summarized in Table~\ref{table:signalyields} for selected mass points.
Tables~\ref{table:cutop_Low}--\ref{table:cutop_High_em} show the optimized selections applied on top of the low- and high-mass SRs requirements for each mass hypothesis.
These tables also present the observed event counts in data and the expected background for each signal mass hypothesis.
The data are generally consistent with the predicted background in all three flavor channels.
The largest deviation observed is in the $\mu\mu$ channel of SR1, at a signal mass of 600\GeV, and has a local significance of 2.3 standard deviations.
The corresponding point of SR2 does not show a matching fluctuation.

\begin{table*}[tb]
\topcaption{
Numbers of expected signal events passing the selection requirements.
The matrix element squared is assumed to be
$1 \times 10^{-4}$,
$1 \times 10^{-2}$, and
1 for $\mN = 40$, 200, and 1000\GeV, respectively.
}
\label{table:signalyields}
\centering
\begin{tabular}{rrrrrrr}
\hline
\mN      &  \multicolumn{2}{c}{$\Pe \Pe$ channel}  & \multicolumn{2}{c}{$\mu \mu$ channel}& \multicolumn{2}{c}{ $\Pe \mu$ channel} \\
(\GeVns)                     &  SR1      & SR2       &  SR1    & SR2 &  SR1    & SR2\\
\hline
40  & 18  & 30  & 33  & 83  & 19  & 42 \\
200  & 5.5  & 0.74  & 9.7  & 1.9  & 7.0  & 1.1 \\
1000  & 0.43  & 4.0  & 0.80  & 7.5  & 0.57  & 4.5 \\
\hline
\end{tabular}
\end{table*}

\begin{figure}[!hptb]
\centering
    \includegraphics[width=0.6\linewidth]{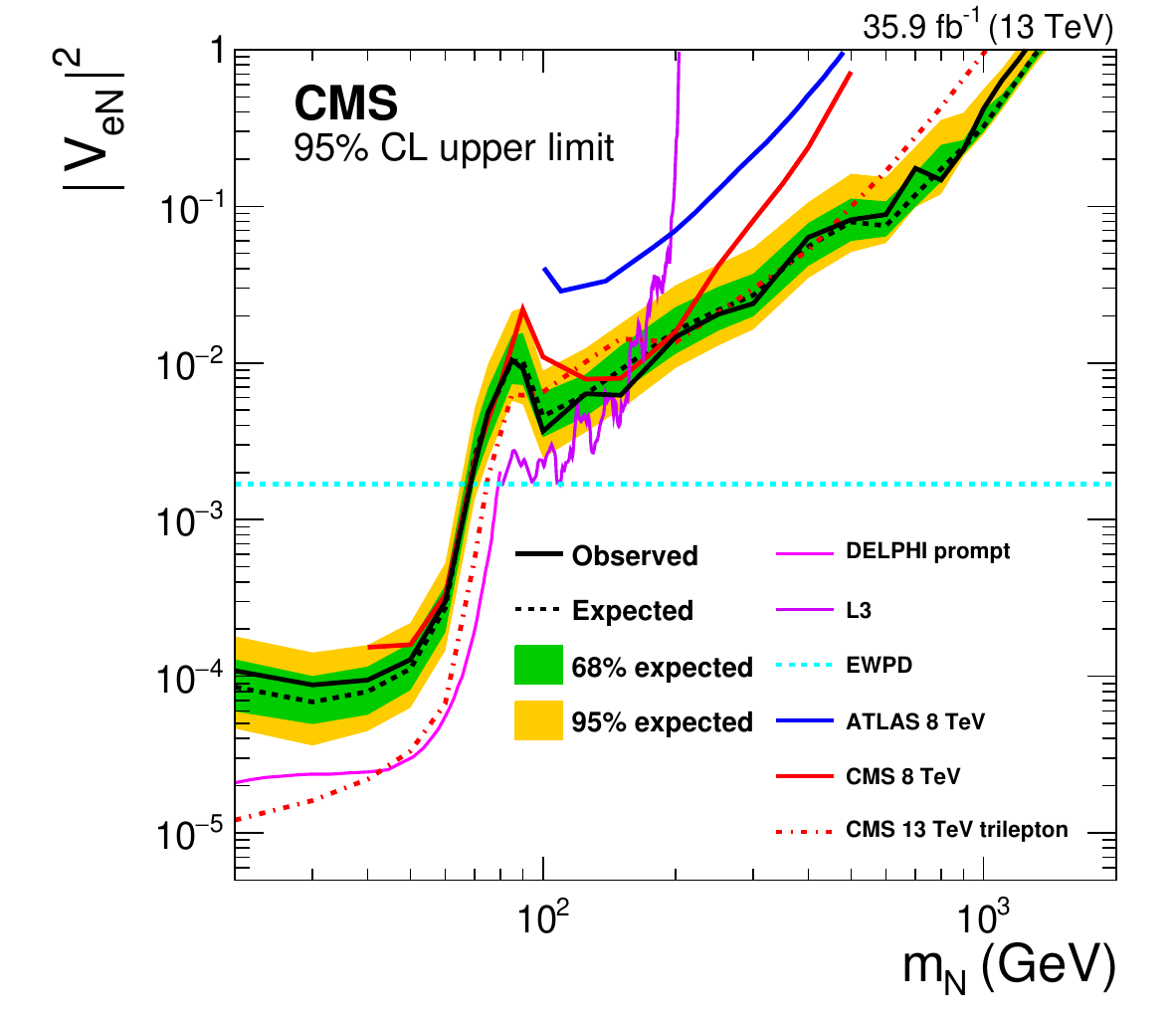}
    \includegraphics[width=0.6\linewidth]{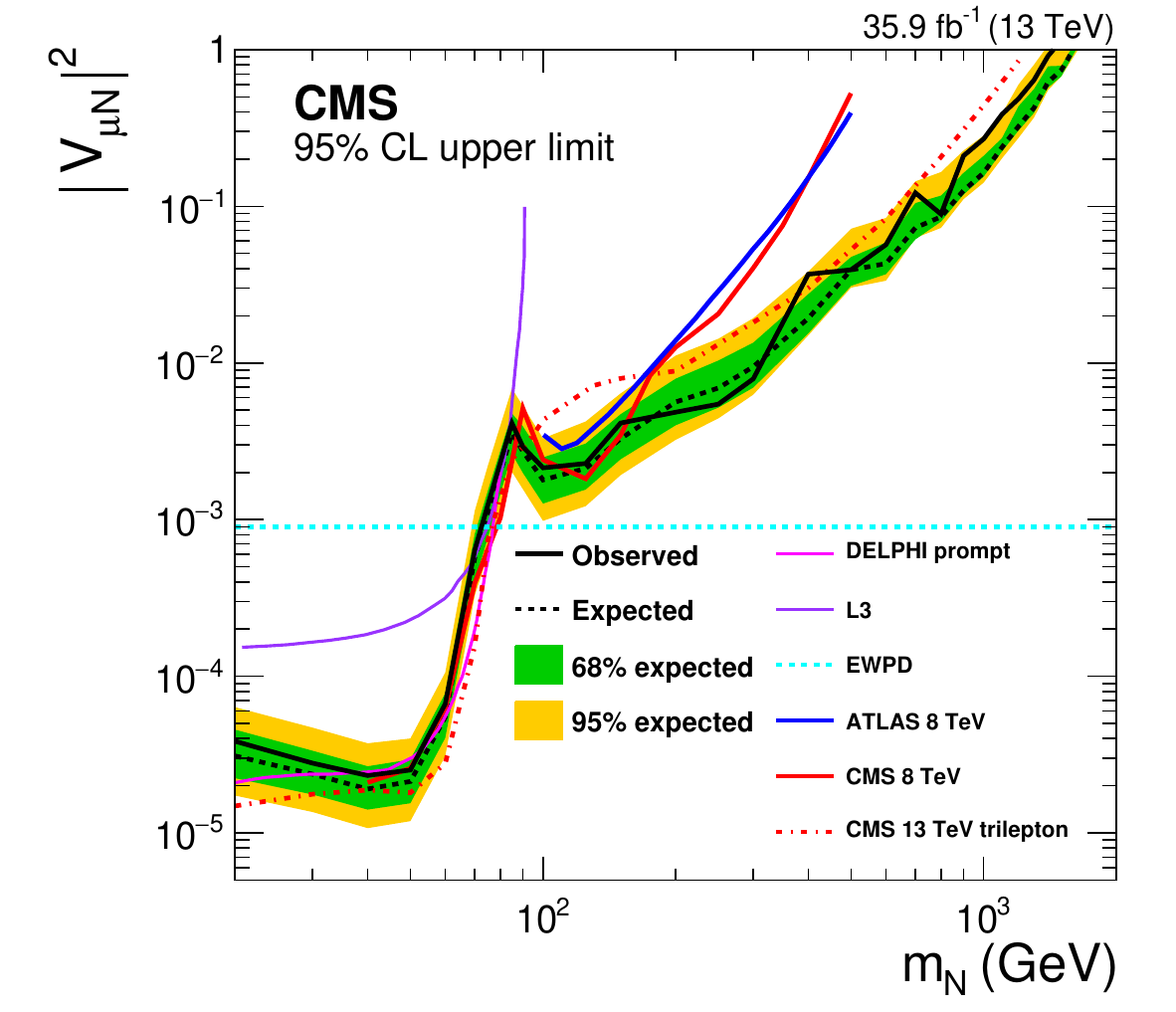}
    \caption{
  Exclusion region at 95\% \CL in the \VeNsq (upper) and \VmNsq (lower) vs. \mN plane.
  The dashed black curve is the expected upper limit, with one and two standard-deviation bands shown in green and yellow, respectively.
  The solid black curve is the observed upper limit.
  The dashed cyan line shows constraints from EWPD~\cite{2013EPJWC..6019008D}.
Also shown are the upper limits from other direct searches:  DELPHI~\cite{delphi}, L3~\cite{l3, l3_2001}, ATLAS~\cite{ATLAS_NR_2012}, and the upper limits from the CMS $\sqrt{s} = 8\TeV$ 2012 data~\cite{CMS_NR_emu_2012} and the trilepton analysis~\cite{Sirunyan:2018mtv} based on the same 2016 data set as used in this analysis.}
    \label{fig:limit_ElElMuMu}
\end{figure}

\begin{figure}[!hptb]
\centering
    \includegraphics[width=0.6\textwidth]{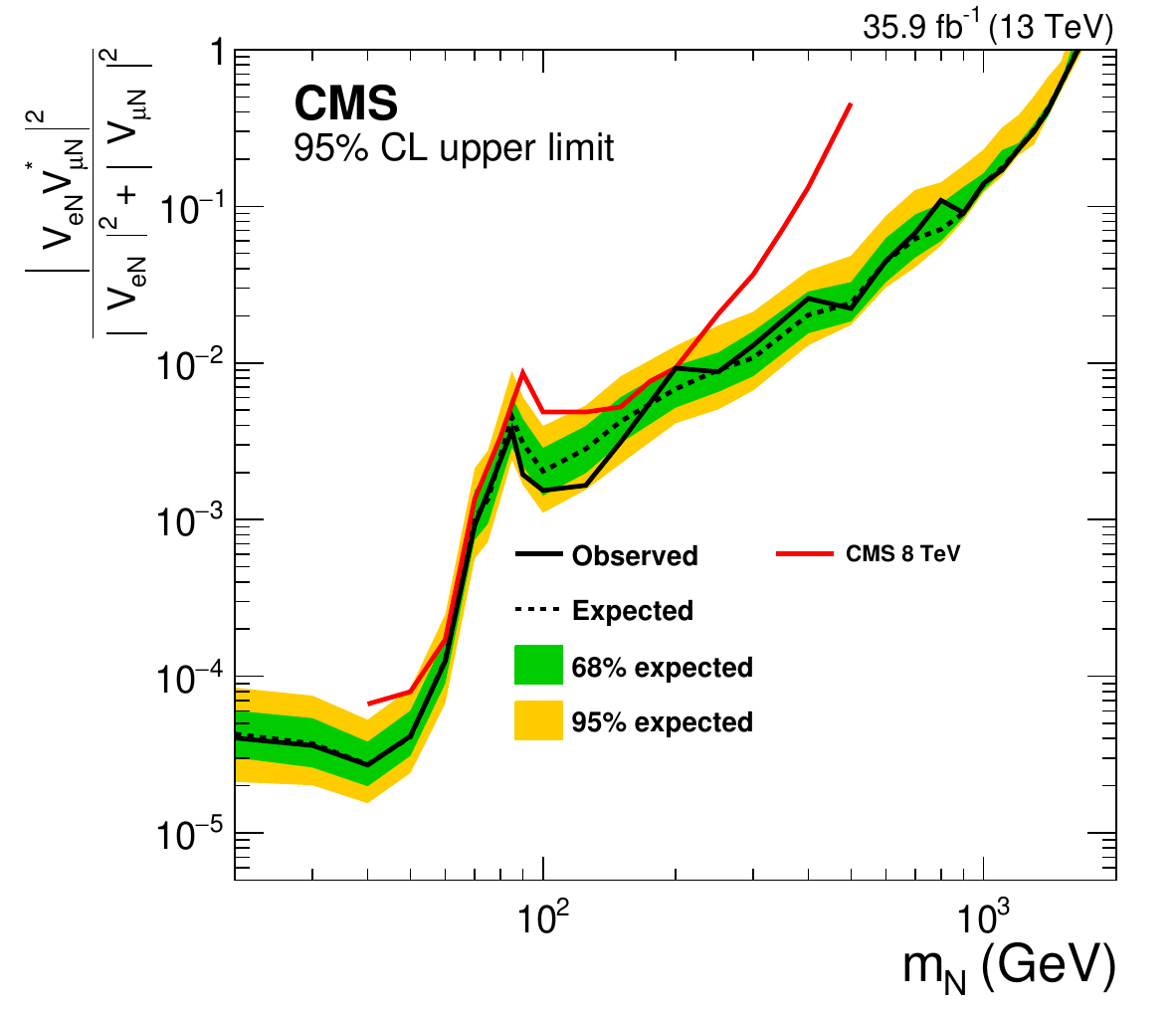}
    \caption{
Exclusion region at 95\% \CL in the \VemNsq vs. \mN plane.
  The dashed black curve is the expected upper limit, with one and two standard-deviation bands shown in green and yellow, respectively.
  The solid black curve is the observed upper limit.
Also shown are the upper limits from the CMS $\sqrt{s} = 8\TeV$ 2012 data~\cite{CMS_NR_emu_2012}.
    }
    \label{fig:limit_MuEl}
\end{figure}

Exclusion limits at 95\% confidence level (\CL) are set on the heavy Majorana neutrino mixing matrix elements as a function of \mN.
The limits are obtained using \CLs criterion~\cite{cls2,cls} based on the event yields in Tables~\ref{table:cutop_Low}--\ref{table:cutop_High_em}.
Log-normal distributions are used for both the signal and nuisance parameters.
The combined limits from SR1 and SR2, on the absolute values of the matrix elements \VeNsq, \VmNsq, and \VemNsq
are shown in Figs.~\ref{fig:limit_ElElMuMu}--\ref{fig:limit_MuEl}, also as a function of \mN.
We assume the systematic uncertainties in SR1 and SR2 to be fully correlated when calculating these limits.
The limits are calculated separately for each of the three channels.
For an \N mass of 40\GeV the observed (expected) limits are
$\abs{\VeN}^{2} < 9.5\,(8.0) \times 10^{-5}$,
$\abs{\VmN}^{2} < 2.3\,(1.9) \times 10^{-5}$,
and $\VemNsq < 2.7\,(2.7) \times 10^{-5}$,
and for an \N mass of 1000\GeV the limits are
$\abs{\VeN}^{2} < 0.42\,(0.32)$,
$\abs{\VmN}^{2} < 0.27\,(0.16)$,
and $\VemNsq < 0.14\,(0.14)$.

The mass range below $\mN = 20\GeV$ is not considered because of the very low selection efficiency in this region.
Furthermore, since the \N lifetime is inversely proportional to $\mN^5 \VlNsq$, for $\mN < 20\GeV$ it becomes significant and results in displaced decays.
Thus the prompt lepton requirement is not satisfied.
The behavior of the limits around $\mN = 80\GeV$ is caused by the fact that as the mass of the heavy Majorana neutrino approaches the $\PW$ boson mass, the
lepton produced together with the \N or the lepton from the \N decay has very low \pt.

The present search at 13\TeV extends the previous CMS SS2$\ell$ plus jets searches at 8\TeV~\cite{CMS_NR_mu_2012,CMS_NR_emu_2012} to both higher \N masses as well as lower masses.
In those earlier searches, two AK4 jets were required in the low- and high-mass SRs,
while in the present analysis at $\sqrt{s} = 13\TeV$, the search has been extended in the low-mass SR to include events with exactly one AK4 jet, and in the high-mass SR to include events with at least one AK8 jet.
As seen in Figs.~\ref{fig:limit_ElElMuMu}--\ref{fig:limit_MuEl}, the exclusion limits for the mixing matrix elements are extended both for low and high \N mass, and now cover \N masses from 20 to 1600\GeV.
In the range previously studied, the present limits significantly improve over the previous results except in the region from 60--80\GeV, where they are equivalent.
The 13\TeV data were taken at higher collision rates and thus with higher trigger thresholds and pileup rates, which impacted the sensitivity of the search in the low-mass region.
This region is covered with high efficiency by a recent search in trilepton channels~\cite{Sirunyan:2018mtv}.

Fig.~\ref{fig:limit_ElElMuMu} shows the exclusion limits for \VeNsq and \VmNsq overlaid with the 13\TeV CMS limits from the trilepton channel~\cite{Sirunyan:2018mtv} and the limits from LEP~\cite{delphi,l3, l3_2001}.
The LEP analyses search for $s$- and $t$-channel production of \N in the process $\Pe\Pe \to \N_{\ell}\nu$, where $\ell$ denotes $\Pe$ or $\mu$.
The contribution of the $t$-channel process (which is only possible in the electron channel) to the total cross section is dominant, and as a result for masses above the $\PW$ boson mass LEP is not sensitive to the muon channel.
The experimental conditions at LEP allow for a low-background search, with high signal efficiency, and as a consequence the results from DELPHI are particularly strong for neutrino masses below the $\PW$ boson mass for both \VeNsq and \VmNsq, while L3 sets strong limits on \VeNsq for masses in the range 80--205\GeV.
For low-mass signals the trilepton analysis is more sensitive, since it has both a smaller level of background from misidentified leptons and higher signal efficiency.
The efficiencies for high-mass signals are comparable, however with the inclusion of the additional SR (using AK8 jets) and the larger branching fraction in the dilepton channel, this analysis has more stringent limits for \N masses above 100\GeV.

\begin{table}[htbp]
\topcaption{
Selection requirements on discriminating variables determined by the optimization
for each Majorana neutrino mass point in the low-mass signal regions.
Columns 8 and 9 show the total background yields (Total bkgd.) and the number of observed data (\nobs), respectively.
The last column shows the overall signal acceptances for the DY channel.
The quoted uncertainties include both the statistical and systematic contributions.}
\label{table:cutop_Low}
  \centering
  \cmsTable{
  \begin{tabular}{ccccccccccc}
\hline
\mN & $p^{\ell_1}_{\mathrm{T}}$  & $\pt^{\ell_2}$ & $m(\ell^{\pm} \ell^{\pm} \pwjet)$  & $m(\ell_{1} \pwjet)$& $m(\ell_{2} \pwjet)$ & $m(\ell^{\pm} \ell^{\pm})$ & Total bkgd. & \nobs & DY $A\epsilon$ \\
(\GeVns) & (\GeVns) &  (\GeVns)  & (\GeVns) & (\GeVns) & (\GeVns) &(\GeVns) & & & (\%) \\
\hline
\textbf{\boldmath{$\Pe\Pe$} channel SR1} &&&&&&&& \\
20 & 25--70 & 60 & $<$190 & $<$160 & $<$160 & 10--60 & $\phantom{}48.9 \pm 9.5\phantom{0}$ & 45 & $0.12\pm 0.02$ \\
30 & 25--70 & 60 & $<$190 & $<$160 & $<$160 & 10--60 & $\phantom{}48.9 \pm 9.5\phantom{0}$ & 45 & $0.13\pm 0.02$ \\
40 & 25--70 & 60 & $<$190 & $<$160 & $<$160 & 10--60 & $\phantom{}48.9 \pm 9.5\phantom{0}$ & 45 & $0.21\pm 0.03$ \\
50 & 25--70 & 60 & $<$190 & $<$160 & $<$160 & 10--60 & $\phantom{}48.9 \pm 9.5\phantom{0}$ & 45 & $0.24\pm 0.03$ \\
60 & 25--70 & 60 & $<$190 & $<$160 & $<$160 & 10--60 & $\phantom{}48.9 \pm 9.5\phantom{0}$ & 45 & $0.18\pm 0.02$ \\
70 & 25--70 & 60 & $<$190 & $<$160 & $<$160 & 10--75 & $\phantom{.0}64 \pm 12\phantom{.0}$ & 58 & $0.10\pm 0.01$ \\
75 & 25--70 & 60 & $<$190 & $<$160 & $<$160 & 10--100 & $\phantom{.0}68 \pm 12\phantom{.0}$ & 67 & $0.13\pm 0.02$ \\
\textbf{\boldmath{$\Pe\Pe$} channel SR2} &&&&&&&& \\
20 & 25--70 & 60 & $<$100 & $<$70 & $<$70 & 10--60 & $\phantom{}50.3 \pm 8.5\phantom{0}$ & 55 & $0.26\pm 0.03$ \\
30 & 25--70 & 60 & $<$100 & $<$70 & $<$70 & 10--60 & $\phantom{}50.3 \pm 8.5\phantom{0}$ & 55 & $0.30\pm 0.04$ \\
40 & 25--70 & 60 & $<$100 & $<$70 & $<$70 & 10--60 & $\phantom{}50.3 \pm 8.5\phantom{0}$ & 55 & $0.35\pm 0.04$ \\
50 & 25--70 & 60 & $<$100 & $<$70 & $<$70 & 10--60 & $\phantom{}50.3 \pm 8.5\phantom{0}$ & 55 & $0.32\pm 0.03$ \\
60 & 25--70 & 60 & $<$100 & $<$70 & $<$70 & 10--60 & $\phantom{}50.3 \pm 8.5\phantom{0}$ & 55 & $0.24\pm 0.03$ \\
70 & 25--70 & 60 & $<$100 & $<$70 & $<$70 & 10--75 & $\phantom{.0}65 \pm 10\phantom{.0}$ & 70 & $0.06\pm 0.01$ \\
75 & 25--70 & 60 & $<$100 & $<$70 & $<$70 & 10--80 & $\phantom{.0}67 \pm 10\phantom{.0}$ & 70 & $0.11\pm 0.02$ \\
\textbf{\boldmath{$\mu\mu$} channel SR1} &&&&&&&& \\
20 & 20--80 & 15--50 & $<$160 & $<$150 & $<$150 & 20--60 & $\phantom{}15.3 \pm 3.4\phantom{0}$ & 18 & $0.10\pm 0.02$ \\
30 & 20--80 & 15--50 & $<$160 & $<$150 & $<$150 & 20--60 & $\phantom{}15.3 \pm 3.4\phantom{0}$ & 18 & $0.18\pm 0.03$ \\
40 & 20--80 & 15--50 & $<$160 & $<$150 & $<$150 & 20--60 & $\phantom{}15.3 \pm 3.4\phantom{0}$ & 18 & $0.34\pm 0.05$ \\
50 & 20--80 & 15--50 & $<$160 & $<$150 & $<$150 & 20--60 & $\phantom{}15.3 \pm 3.4\phantom{0}$ & 18 & $0.40\pm 0.04$ \\
60 & 20--80 & 15--50 & $<$160 & $<$150 & $<$150 & 20--60 & $\phantom{}15.3 \pm 3.4\phantom{0}$ & 18 & $0.33\pm 0.04$ \\
70 & 20--80 & 15--50 & $<$160 & $<$150 & $<$150 & 10--75 & $\phantom{}20.3 \pm 4.4\phantom{0}$ & 21 & $0.17\pm 0.02$ \\
75 & 20--80 & 15--50 & $<$160 & $<$150 & $<$150 & 20--100 & $\phantom{}18.9 \pm 4.0\phantom{0}$ & 19 & $0.19\pm 0.03$ \\
\textbf{\boldmath{$\mu\mu$} channel SR2} &&&&&&&& \\
20 & 20--80 & 15--50 & $<$100 & $<$70 & $<$70 & 20--60 & $\phantom{}25.9 \pm 5.9\phantom{0}$ & 29 & $0.28\pm 0.03$ \\
30 & 20--80 & 15--50 & $<$100 & $<$70 & $<$70 & 20--60 & $\phantom{}25.9 \pm 5.9\phantom{0}$ & 29 & $0.51\pm 0.05$ \\
40 & 20--80 & 15--50 & $<$100 & $<$70 & $<$70 & 20--60 & $\phantom{}25.9 \pm 5.9\phantom{0}$ & 29 & $0.8\pm 0.1$ \\
50 & 20--80 & 15--50 & $<$100 & $<$70 & $<$70 & 20--60 & $\phantom{}25.9 \pm 5.9\phantom{0}$ & 29 & $1.1\pm 0.1$ \\
60 & 20--80 & 15--50 & $<$100 & $<$70 & $<$70 & 20--60 & $\phantom{}25.9 \pm 5.9\phantom{0}$ & 29 & $0.73\pm 0.07$ \\
70 & 20--80 & 15--50 & $<$100 & $<$70 & $<$70 & 10--75 & $\phantom{}37.5 \pm 7.1\phantom{0}$ & 41 & $0.20\pm 0.03$ \\
75 & 20--80 & 15--50 & $<$100 & $<$70 & $<$70 & 20--80 & $\phantom{}29.7 \pm 6.7\phantom{0}$ & 34 & $0.24\pm 0.03$ \\
\textbf{\boldmath{$\Pe\mu$} channel SR1} &&&&&&&& \\
20 & 25--60 & 15--40 & $<$185 & $<$135 & $<$135 & 20--60 & $\phantom{}34.0 \pm 6.4\phantom{0}$ & 34 & $0.08\pm 0.02$ \\
30 & 25--60 & 15--40 & $<$185 & $<$135 & $<$135 & 20--60 & $\phantom{}34.0 \pm 6.4\phantom{0}$ & 34 & $0.12\pm 0.02$ \\
40 & 25--60 & 15--40 & $<$185 & $<$135 & $<$135 & 20--60 & $\phantom{}34.0 \pm 6.4\phantom{0}$ & 34 & $0.21\pm 0.02$ \\
50 & 25--60 & 15--40 & $<$185 & $<$135 & $<$135 & 20--60 & $\phantom{}34.0 \pm 6.4\phantom{0}$ & 34 & $0.20\pm 0.03$ \\
60 & 25--60 & 15--40 & $<$185 & $<$135 & $<$135 & 20--60 & $\phantom{}34.0 \pm 6.4\phantom{0}$ & 34 & $0.17\pm 0.02$ \\
70 & 25--60 & 15--40 & $<$185 & $<$135 & $<$135 & 10--75 & $\phantom{.0}51 \pm 10\phantom{.0}$ & 49 & $0.09\pm 0.01$ \\
75 & 25--60 & 15--40 & $<$185 & $<$135 & $<$135 & 20--100 & $\phantom{}46.5 \pm 8.7\phantom{0}$ & 49 & $0.17\pm 0.03$ \\
\textbf{\boldmath{$\Pe\mu$} channel SR2} &&&&&&&& \\
20 & 25--60 & 15--40 & $<$100 & $<$65 & $<$65 & 20--60 & $\phantom{}51.7 \pm 9.2\phantom{0}$ & 50 & $0.21\pm 0.02$ \\
30 & 25--60 & 15--40 & $<$100 & $<$65 & $<$65 & 20--60 & $\phantom{}51.7 \pm 9.2\phantom{0}$ & 50 & $0.27\pm 0.03$ \\
40 & 25--60 & 15--40 & $<$100 & $<$65 & $<$65 & 20--60 & $\phantom{}51.7 \pm 9.2\phantom{0}$ & 50 & $0.45\pm 0.04$ \\
50 & 25--60 & 15--40 & $<$100 & $<$65 & $<$65 & 20--60 & $\phantom{}51.7 \pm 9.2\phantom{0}$ & 50 & $0.40\pm 0.03$ \\
60 & 25--60 & 15--40 & $<$100 & $<$65 & $<$65 & 20--60 & $\phantom{}51.7 \pm 9.2\phantom{0}$ & 50 & $0.24\pm 0.03$ \\
70 & 25--60 & 15--40 & $<$100 & $<$65 & $<$65 & 10--75 & $\phantom{}75.8 \pm 12.4\phantom{}$ & 65 & $0.09\pm 0.01$ \\
75 & 25--60 & 15--40 & $<$100 & $<$65 & $<$65 & 20--80 & $\phantom{}62.8 \pm 10.9\phantom{}$ & 57 & $0.12\pm 0.03$ \\
\hline
\end{tabular}
}
\end{table}

\begin{table}[!hptb]
  \centering
  \topcaption{
  Selection requirements on discriminating variables determined by the optimization for each Majorana neutrino mass point in the $\Pe\Pe$ channel high-mass SRs. 
  Columns 7 and 8 show the total background yields (Total bkgd.) and the number of observed data (\nobs), respectively.
  The last columns show the overall signal acceptance for the DY and VBF channels.
  The quoted uncertainties include both the statistical and systematic contributions.
  The dash indicates that no selection requirement is made.
  }
  \label{table:cutop_High_ee}
  \cmsTable{
    \begin{tabular}{cccccccccc}
\hline
\mN & $p^{\ell_1}_{\mathrm{T}}$  & $\pt^{\ell_2}$ & $m(\ell^{\pm} \ell^{\pm} \pwjet)$  & $m(\ell \pwjet)$  & \metSQst & Total bkgd. & $\nobs$ & DY $A\epsilon$ & VBF $A\epsilon$\\
(\GeVns) & (\GeVns) & (\GeVns) & (\GeVns) & (\GeVns) & (\GeVns) & & & (\%)& (\%)\\
\hline
\textbf{\boldmath{$\Pe \Pe$} channel SR1} &&&&&&&&\\
85 & $>$25 & $>$15 & $>$110 & 45--95 & $<$6 & $\phantom{0}9.5 \pm 2.8\phantom{0}$ & 9 & $\phantom{0.}0.11\pm 0.02\phantom{.}$ & \NA  \\
90 & $>$25 & $>$15 & $>$110 & 50--100 & $<$6 & $\phantom{}12.5 \pm 3.5\phantom{0}$ & 10 & $\phantom{0.}0.23\pm 0.05\phantom{.}$ & \NA  \\
100 & $>$25 & $>$15 & $>$120 & 50--110 & $<$6 & $\phantom{}20.3 \pm 5.0\phantom{0}$ & 15 & $\phantom{0.0}1.1\pm 0.1\phantom{.0}$ & \NA  \\
125 & $>$30 & $>$25 & $>$120 & 90--140 & $<$6 & $\phantom{}17.7 \pm 4.5\phantom{0}$ & 17 & $\phantom{0.0}2.6\pm 0.2\phantom{.0}$ & \NA  \\
150 & $>$40 & $>$25 & $>$180 & 130--160 & $<$6 & $\phantom{}14.7 \pm 3.8\phantom{0}$ & 9 & $\phantom{0.0}3.1\pm 0.2\phantom{.0}$ & \NA  \\
200 & $>$55 & $>$40 & $>$220 & 160--225 & $<$6 & $\phantom{}12.4 \pm 2.7\phantom{0}$ & 10 & $\phantom{0.0}4.9\pm 0.4\phantom{.0}$ & \NA  \\
250 & $>$70 & $>$60 & $>$310 & 220--270 & $<$6 & $\phantom{0}6.0 \pm 1.7\phantom{0}$ & 4 & $\phantom{0.0}5.9\pm 0.4\phantom{.0}$ & \NA  \\
300 & $>$80 & $>$60 & $>$370 & 235--335 & $<$6 & $\phantom{0}8.2 \pm 2.1\phantom{0}$ & 6 & $\phantom{0.0}7.6\pm 0.5\phantom{.0}$ & $\phantom{.0}3.0\pm 0.3\phantom{.0}$ \\
400 & $>$100 & $>$65 & $>$450 & 335--450 & $<$6 & $\phantom{0}2.5 \pm 1.4\phantom{0}$ & 4 & $\phantom{0.0}6.6\pm 0.5\phantom{.0}$ & $\phantom{.0}3.0\pm 0.2\phantom{.0}$ \\
500 & $>$125 & $>$65 & $>$560 & 400--555 & $<$6 & $\phantom{0}1.5 \pm 0.8\phantom{0}$ & 5 & $\phantom{0.0}5.5\pm 0.4\phantom{.0}$ & $\phantom{.0}2.7\pm 0.2\phantom{.0}$ \\
600 & $>$125 & \NA & $>$760 & 400--690 & $<$6 & $\phantom{0}0.9 \pm 0.6\phantom{0}$ & 1 & $\phantom{0.0}3.8\pm 0.3\phantom{.0}$ & $\phantom{.0}1.7\pm 0.2\phantom{.0}$ \\
700 & $>$125 & \NA & $>$760 & 400--955 & $<$6 & $\phantom{0}1.7 \pm 0.7\phantom{0}$ & 1 & $\phantom{0.0}4.0\pm 0.3\phantom{.0}$ & $\phantom{.0}2.8\pm 0.2\phantom{.0}$ \\
800 & $>$125 & \NA & $>$760 & 400--1130 & $<$6 & $\phantom{0}1.7 \pm 0.7\phantom{0}$ & 1 & $\phantom{0.0}3.6\pm 0.3\phantom{.0}$ & $\phantom{.0}3.0\pm 0.3\phantom{.0}$ \\
900 & $>$125 & \NA & $>$760 & 400--1300 & $<$6 & $\phantom{0}1.7 \pm 0.7\phantom{0}$ & 1 & $\phantom{0.0}3.2\pm 0.2\phantom{.0}$ & $\phantom{.0}2.9\pm 0.2\phantom{.0}$ \\
1000 & $>$125 & \NA & $>$760 & 400--1490 & $<$6 & $\phantom{0}1.7 \pm 0.7\phantom{0}$ & 1 & $\phantom{0.0}2.6\pm 0.2\phantom{.0}$ & $\phantom{.0}2.4\pm 0.2\phantom{.0}$ \\
1100 & $>$125 & \NA & $>$760 & 400--1490 & $<$6 & $\phantom{0}1.7 \pm 0.7\phantom{0}$ & 1 & $\phantom{0.0}2.2\pm 0.2\phantom{.0}$ & $\phantom{.0}2.0\pm 0.2\phantom{.0}$ \\
1200 & $>$125 & \NA & $>$760 & 400--1600 & $<$6 & $\phantom{0}1.7 \pm 0.7\phantom{0}$ & 1 & $\phantom{0.0}2.0\pm 0.2\phantom{.0}$ & $\phantom{.0}1.8\pm 0.2\phantom{.0}$ \\
1300 & $>$125 & \NA & $>$760 & 400--1930 & $<$6 & $\phantom{0}1.7 \pm 0.7\phantom{0}$ & 1 & $\phantom{0.0}1.8\pm 0.1\phantom{.0}$ & $\phantom{.0}1.6\pm 0.2\phantom{.0}$ \\
1400 & $>$125 & \NA & $>$760 & 400--1930 & $<$6 & $\phantom{0}1.7 \pm 0.7\phantom{0}$ & 1 & $\phantom{0.0}1.5\pm 0.1\phantom{.0}$ & $\phantom{.0}1.3\pm 0.1\phantom{.0}$ \\
1500 & $>$125 & \NA & $>$760 & 400--1930 & $<$6 & $\phantom{0}1.7 \pm 0.7\phantom{0}$ & 1 & $\phantom{0.0}1.3\pm 0.1\phantom{.0}$ & $\phantom{.0}1.2\pm 0.2\phantom{.0}$ \\
\textbf{\boldmath{$\Pe \Pe$} channel SR2} &&&&&&\\
85 & $>$25 & $>$15 & \NA &  \NA & $<$15 &  $\phantom{}10.9 \pm 2.9\phantom{0}$ & 10 & $\phantom{0.}0.001\pm 0.001\phantom{.}$ & \NA  \\
90 & $>$25 & $>$15 & \NA &  90--220 & $<$15 &  $\phantom{0}3.4 \pm 1.0\phantom{0}$ & 2 & $\phantom{0.}0.003\pm 0.002\phantom{.}$ & \NA  \\
100 & $>$25 & $>$15 & \NA &  100--220 & $<$15 &  $\phantom{0}3.4 \pm 1.0\phantom{0}$ & 2 & $\phantom{0.}0.005\pm 0.003\phantom{.}$ & \NA  \\
125 & $>$60 & $>$15 & \NA &  123--145 & $<$15 &  $\phantom{0}0.2 \pm 0.1\phantom{0}$ & 0 & $\phantom{0.}0.04\pm 0.01\phantom{.}$ & \NA  \\
150 & $>$90 & $>$15 & \NA &  125--185 & $<$15 &  $\phantom{0}1.3 \pm 0.5\phantom{0}$ & 0 & $\phantom{0.}0.19\pm 0.03\phantom{.}$ & \NA  \\
200 & $>$100 & $>$20 & \NA &  173--220 & $<$15 &  $\phantom{0}0.8 \pm 0.3\phantom{0}$ & 1 & $\phantom{0.}0.60\pm 0.07\phantom{.}$ & \NA  \\
250 & $>$100 & $>$25 & \NA &  220--305 & $<$15 &  $\phantom{0}2.1 \pm 1.2\phantom{0}$ & 3 & $\phantom{0.0}2.2\pm 0.2\phantom{.0}$ & \NA  \\
300 & $>$100 & $>$30 & \NA &  270--330 & $<$15 &  $\phantom{0}1.3 \pm 0.6\phantom{0}$ & 1 & $\phantom{0.0}3.5\pm 0.4\phantom{.0}$ & $\phantom{0.0}0.6\pm 0.1\phantom{.0}$ \\
400 & $>$100 & $>$35 & \NA &  330--440 & $<$15 &  $\phantom{0}3.1 \pm 1.3\phantom{0}$ & 3 & $\phantom{0.0}9.1\pm 0.9\phantom{.0}$ & $\phantom{0.0}2.9\pm 0.3\phantom{.0}$ \\
500 & $>$120 & $>$35 & \NA &  440--565 & $<$15 &  $\phantom{0}2.8 \pm 1.0\phantom{0}$ & 1 & $\phantom{.0}14.3\pm 1.4\phantom{.0}$ & $\phantom{0.0}6.1\pm 0.6\phantom{.0}$ \\
600 & $>$120 & \NA & \NA &  565--675 & $<$15 &  $\phantom{0}0.8 \pm 0.3\phantom{0}$ & 1 & $\phantom{.0}17.4\pm 1.8\phantom{.0}$ & $\phantom{.0}11.0\pm 1.0\phantom{.0}$ \\
700 & $>$140 & \NA & \NA &  635--775 & $<$15 &  $\phantom{0}0.8 \pm 0.3\phantom{0}$ & 2 & $\phantom{.0}19.4\pm 2.0\phantom{.0}$ & $\phantom{.0}13.1\pm 1.3\phantom{.0}$ \\
800 & $>$140 & \NA & \NA &  740--1005 & $<$15 &  $\phantom{0}0.9 \pm 0.4\phantom{0}$ & 0 & $\phantom{.0}20.8\pm 2.1\phantom{.0}$ & $\phantom{.0}14.0\pm 1.3\phantom{.0}$ \\
900 & $>$140 & \NA & \NA &  865--1030 & $<$15 &  $\phantom{0}0.2 \pm 0.1\phantom{0}$ & 0 & $\phantom{.0}19.2\pm 2.0\phantom{.0}$ & $\phantom{.0}13.2\pm 1.3\phantom{.0}$ \\
1000 & $>$140 & \NA & \NA &  890--1185 & $<$15 &  $\phantom{0}0.3 \pm 0.1\phantom{0}$ & 1 & $\phantom{.0}21.5\pm 2.2\phantom{.0}$ & $\phantom{.0}15.3\pm 1.5\phantom{.0}$ \\
1100 & $>$140 & \NA & \NA &  1035--1395 & $<$15 &  $\phantom{0}0.1 \pm 0.1\phantom{0}$ & 1 & $\phantom{.0}20.3\pm 2.1\phantom{.0}$ & $\phantom{.0}14.7\pm 1.4\phantom{.0}$ \\
1200 & $>$140 & \NA & \NA &  1085--1460 & $<$15 &  $\phantom{0}0.1 \pm 0.0\phantom{0}$ & 1 & $\phantom{.0}20.8\pm 2.2\phantom{.0}$ & $\phantom{.0}15.3\pm 1.5\phantom{.0}$ \\
1300 & $>$140 & \NA & \NA &  1140--1590 & $<$15 &  $\phantom{0}0.1 \pm 0.0\phantom{0}$ & 1 & $\phantom{.0}20.5\pm 2.2\phantom{.0}$ & $\phantom{.0}15.5\pm 1.6\phantom{.0}$ \\
1400 & $>$140 & \NA & \NA &  1245--1700 & $<$15 &  $\phantom{0}0.1 \pm 0.0\phantom{0}$ & 0 & $\phantom{.0}19.6\pm 2.1\phantom{.0}$ & $\phantom{.0}15.1\pm 1.6\phantom{.0}$ \\
1500 & $>$140 & \NA & \NA &  1300--1800 & $<$15 &  $\phantom{0}0.04 \pm 0.02\phantom{0}$ & 0 & $\phantom{.0}19.5\pm 2.1\phantom{.0}$ & $\phantom{.0}15.2\pm 1.6\phantom{.0}$ \\
\hline
    \end{tabular}
  }
\end{table}

\begin{table}[!hptb]
  \centering
  \topcaption{
  Selection requirements on discriminating variables determined by the optimization for each Majorana neutrino mass point in the $\mu\mu$ channel high-mass SRs. 
  Columns 7 and 8 show the total background yields (Total bkgd.) and the number of observed data (\nobs), respectively.
  The last columns show the overall signal acceptance for the DY and VBF channels.
  The quoted uncertainties include both the statistical and systematic contributions.
  The dash indicates that no selection requirement is made.
  }
  \label{table:cutop_High_mm}
  \cmsTable{
    \begin{tabular}{cccccccccc}
\hline
\mN & $p^{\ell_1}_{\mathrm{T}}$  & $\pt^{\ell_2}$ & $m(\ell^{\pm} \ell^{\pm} \pwjet)$  & $m(\ell \pwjet)$  & \metSQst & Total bkgd. & $\nobs$ & DY $A\epsilon$ & VBF $A\epsilon$\\
(\GeVns) & (\GeVns) & (\GeVns) & (\GeVns) & (\GeVns) & (\GeVns) & & & (\%)& (\%)\\
\hline
\textbf{\boldmath{$\mu\mu$} channel SR1} &&&&&&&&\\
85 & $>$25 & $>$10 & $>$90 & 40--100 & $<$9 & $\phantom{}26.0 \pm 6.3\phantom{0}$ & 30 & $\phantom{0.}0.50\pm 0.05\phantom{.}$ & \NA  \\
90 & $>$25 & $>$10 & $>$90 & 45--105 & $<$9 & $\phantom{}34.5 \pm 7.5\phantom{0}$ & 35 & $\phantom{0.0}1.2\pm 0.1\phantom{.0}$ & \NA  \\
100 & $>$25 & $>$15 & $>$110 & 55--115 & $<$9 & $\phantom{}18.6 \pm 4.2\phantom{0}$ & 20 & $\phantom{0.0}2.6\pm 0.2\phantom{.0}$ & \NA  \\
125 & $>$25 & $>$25 & $>$140 & 85--140 & $<$7 & $\phantom{}11.7 \pm 2.7\phantom{0}$ & 12 & $\phantom{0.0}5.1\pm 0.4\phantom{.0}$ & \NA  \\
150 & $>$35 & $>$35 & $>$150 & 110--170 & $<$7 & $\phantom{0}8.9 \pm 1.9\phantom{0}$ & 11 & $\phantom{0.0}6.6\pm 0.5\phantom{.0}$ & \NA  \\
200 & $>$50 & $>$40 & $>$250 & 160--215 & $<$7 & $\phantom{0}4.6 \pm 1.2\phantom{0}$ & 4 & $\phantom{0.0}8.1\pm 0.6\phantom{.0}$ & \NA  \\
250 & $>$85 & $>$45 & $>$310 & 215--270 & $<$7 & $\phantom{0}3.0 \pm 0.9\phantom{0}$ & 2 & $\phantom{.0}11.0\pm 0.8\phantom{.0}$ & \NA  \\
300 & $>$100 & $>$50 & $>$370 & 225--340 & $<$7 & $\phantom{0}2.6 \pm 1.0\phantom{0}$ & 2 & $\phantom{.0}13.2\pm 0.9\phantom{.0}$ & $\phantom{.0}5.2\pm 0.4\phantom{.0}$ \\
400 & $>$110 & $>$60 & $>$490 & 295--490 & $<$7 & $\phantom{0}0.9 \pm 0.4\phantom{0}$ & 3 & $\phantom{.0}11.7\pm 0.8\phantom{.0}$ & $\phantom{.0}5.1\pm 0.4\phantom{.0}$ \\
500 & $>$110 & $>$60 & $>$610 & 370--550 & $<$7 & $\phantom{0}0.4^{\phantom{0}+\phantom{0}0.6\phantom{0}\phantom{0}}_{\phantom{0}-\phantom{0}0.4\phantom{0}\phantom{0}}$ & 3 & $\phantom{0.0}8.6\pm 0.6\phantom{.0}$ & $\phantom{.0}4.1\pm 0.3\phantom{.0}$ \\
600 & $>$110 & \NA & $>$680 & 370--630 & $<$7 & $\phantom{0}0.3^{\phantom{0}+\phantom{0}0.3\phantom{0}\phantom{0}}_{\phantom{0}-\phantom{0}0.3\phantom{0}\phantom{0}}$ & 3 & $\phantom{0.0}7.4\pm 0.5\phantom{.0}$ & $\phantom{.0}4.1\pm 0.3\phantom{.0}$ \\
700 & $>$110 & \NA & $>$800 & 370--885 & $<$7 & $\phantom{0}0.2^{\phantom{0}+\phantom{0}0.4\phantom{0}\phantom{0}}_{\phantom{0}-\phantom{0}0.2\phantom{0}\phantom{0}}$ & 2 & $\phantom{0.0}6.7\pm 0.4\phantom{.0}$ & $\phantom{.0}3.9\pm 0.3\phantom{.0}$ \\
800 & $>$110 & \NA & $>$800 & 370--890 & $<$7 & $\phantom{0}0.2^{\phantom{0}+\phantom{0}0.4\phantom{0}\phantom{0}}_{\phantom{0}-\phantom{0}0.2\phantom{0}\phantom{0}}$ & 2 & $\phantom{0.0}6.0\pm 0.4\phantom{.0}$ & $\phantom{.0}5.4\pm 0.3\phantom{.0}$ \\
900 & $>$110 & \NA & $>$800 & 370--1225 & $<$7 & $\phantom{0}0.3^{\phantom{0}+\phantom{0}0.4\phantom{0}\phantom{0}}_{\phantom{0}-\phantom{0}0.3\phantom{0}\phantom{0}}$ & 2 & $\phantom{0.0}5.4\pm 0.4\phantom{.0}$ & $\phantom{.0}5.0\pm 0.3\phantom{.0}$ \\
1000 & $>$110 & \NA & $>$800 & 370--1230 & $<$7 & $\phantom{0}0.3^{\phantom{0}+\phantom{0}0.4\phantom{0}\phantom{0}}_{\phantom{0}-\phantom{0}0.3\phantom{0}\phantom{0}}$ & 2 & $\phantom{0.0}4.6\pm 0.3\phantom{.0}$ & $\phantom{.0}4.2\pm 0.3\phantom{.0}$ \\
1100 & $>$110 & \NA & $>$800 & 370--1245 & $<$7 & $\phantom{0}0.3^{\phantom{0}+\phantom{0}0.4\phantom{0}\phantom{0}}_{\phantom{0}-\phantom{0}0.3\phantom{0}\phantom{0}}$ & 2 & $\phantom{0.0}4.1\pm 0.3\phantom{.0}$ & $\phantom{.0}3.8\pm 0.3\phantom{.0}$ \\
1200 & $>$110 & \NA & $>$800 & 370--1690 & $<$7 & $\phantom{0}0.3^{\phantom{0}+\phantom{0}0.4\phantom{0}\phantom{0}}_{\phantom{0}-\phantom{0}0.3\phantom{0}\phantom{0}}$ & 2 & $\phantom{0.0}3.6\pm 0.2\phantom{.0}$ & $\phantom{.0}3.4\pm 0.3\phantom{.0}$ \\
1300 & $>$110 & \NA & $>$800 & 370--1890 & $<$7 & $\phantom{0}0.3^{\phantom{0}+\phantom{0}0.4\phantom{0}\phantom{0}}_{\phantom{0}-\phantom{0}0.3\phantom{0}\phantom{0}}$ & 2 & $\phantom{0.0}3.2\pm 0.2\phantom{.0}$ & $\phantom{.0}3.0\pm 0.2\phantom{.0}$ \\
1400 & $>$110 & \NA & $>$800 & 370--1940 & $<$7 & $\phantom{0}0.3^{\phantom{0}+\phantom{0}0.4\phantom{0}\phantom{0}}_{\phantom{0}-\phantom{0}0.3\phantom{0}\phantom{0}}$ & 2 & $\phantom{0.0}2.7\pm 0.2\phantom{.0}$ & $\phantom{.0}2.7\pm 0.2\phantom{.0}$ \\
1500 & $>$110 & \NA & $>$800 & 370--2220 & $<$7 & $\phantom{0}0.3^{\phantom{0}+\phantom{0}0.4\phantom{0}\phantom{0}}_{\phantom{0}-\phantom{0}0.3\phantom{0}\phantom{0}}$ & 2 & $\phantom{0.0}2.5\pm 0.2\phantom{.0}$ & $\phantom{.0}2.3\pm 0.2\phantom{.0}$ \\
\textbf{\boldmath{$\mu \mu$} channel SR2} &&&&&&\\
85 & $>$25 & $>$10 & \NA &  \NA & $<$15 &  $\phantom{}11.4 \pm 3.5\phantom{0}$ & 13 & $\phantom{0.}0.001\pm 0.001\phantom{.}$ & \NA  \\
90 & $>$25 & $>$10 & \NA &  90--170 & $<$15 &  $\phantom{0}4.1 \pm 1.3\phantom{0}$ & 4 & $\phantom{0.}0.003\pm 0.003\phantom{.}$ & \NA  \\
100 & $>$25 & $>$15 & \NA &  98--145 & $<$15 &  $\phantom{0}1.0 \pm 0.3\phantom{0}$ & 0 & $\phantom{0.}0.006\pm 0.003\phantom{.}$ & \NA  \\
125 & $>$60 & $>$15 & \NA &  110--150 & $<$15 &  $\phantom{0}0.8 \pm 0.3\phantom{0}$ & 0 & $\phantom{0.}0.08\pm 0.01\phantom{.}$ & \NA  \\
150 & $>$70 & $>$15 & \NA &  145--175 & $<$15 &  $\phantom{0}1.0 \pm 0.4\phantom{0}$ & 2 & $\phantom{0.}0.28\pm 0.04\phantom{.}$ & \NA  \\
200 & $>$100 & $>$20 & \NA &  175--235 & $<$15 &  $\phantom{0}1.3 \pm 0.8\phantom{0}$ & 0 & $\phantom{0.0}1.4\pm 0.1\phantom{.0}$ & \NA  \\
250 & $>$140 & $>$25 & \NA &  226--280 & $<$15 &  $\phantom{0}0.3 \pm 0.2\phantom{0}$ & 0 & $\phantom{0.0}3.0\pm 0.3\phantom{.0}$ & \NA \\
300 & $>$140 & $>$40 & \NA &  280--340 & $<$15 &  $\phantom{0}0.4 \pm 0.3\phantom{0}$ & 0 & $\phantom{0.0}5.4\pm 0.5\phantom{.0}$ & $\phantom{0.0}0.7\pm 0.1\phantom{.0}$ \\
400 & $>$140 & $>$65 & \NA &  340--445 & $<$15 &  $\phantom{0}0.5 \pm 0.3\phantom{0}$ & 2 & $\phantom{.0}13.3\pm 1.3\phantom{.0}$ & $\phantom{0.0}2.7\pm 0.3\phantom{.0}$ \\
500 & $>$140 & $>$65 & \NA &  445--560 & $<$15 &  $\phantom{0}0.8 \pm 0.5\phantom{0}$ & 0 & $\phantom{.0}22.4\pm 2.2\phantom{.0}$ & $\phantom{0.0}6.8\pm 0.7\phantom{.0}$ \\
600 & $>$140 & \NA & \NA &  560--685 & $<$15 &  $\phantom{0}0.7 \pm 0.4\phantom{0}$ & 0 & $\phantom{.0}30.2\pm 2.9\phantom{.0}$ & $\phantom{.0}20.4\pm 1.8\phantom{.0}$ \\
700 & $>$140 & \NA & \NA &  635--825 & $<$15 &  $\phantom{0}0.8 \pm 0.4\phantom{0}$ & 2 & $\phantom{.0}34.6\pm 3.4\phantom{.0}$ & $\phantom{.0}24.7\pm 2.2\phantom{.0}$ \\
800 & $>$140 & \NA & \NA &  755--960 & $<$15 &  $\phantom{0}0.4 \pm 0.3\phantom{0}$ & 0 & $\phantom{.0}34.8\pm 3.5\phantom{.0}$ & $\phantom{.0}24.9\pm 2.3\phantom{.0}$ \\
900 & $>$140 & \NA & \NA &  840--1055 & $<$15 &  $\phantom{0}0.2^{\phantom{0}+\phantom{0}0.2\phantom{0}\phantom{0}}_{\phantom{0}-\phantom{0}0.2\phantom{0}\phantom{0}}$ & 1 & $\phantom{.0}35.8\pm 3.6\phantom{.0}$ & $\phantom{.0}26.9\pm 2.5\phantom{.0}$ \\
1000 & $>$140 & \NA & \NA &  900--1205 & $<$15 &  $\phantom{0}0.1^{\phantom{0}+\phantom{0}0.2\phantom{0}\phantom{0}}_{\phantom{0}-\phantom{0}0.1\phantom{0}\phantom{0}}$ & 1 & $\phantom{.0}38.4\pm 3.9\phantom{.0}$ & $\phantom{.0}28.9\pm 2.7\phantom{.0}$ \\
1100 & $>$140 & \NA & \NA &  990--1250 & $<$15 &  $\phantom{0}0.1^{\phantom{0}+\phantom{0}0.2\phantom{0}\phantom{0}}_{\phantom{0}-\phantom{0}0.1\phantom{0}\phantom{0}}$ & 1 & $\phantom{.0}36.7\pm 3.7\phantom{.0}$ & $\phantom{.0}29.2\pm 2.7\phantom{.0}$ \\
1200 & $>$140 & \NA & \NA &  1035--1430 & $<$15 &  $\phantom{0}0.2^{\phantom{0}+\phantom{0}0.3\phantom{0}\phantom{0}}_{\phantom{0}-\phantom{0}0.2\phantom{0}\phantom{0}}$ & 1 & $\phantom{.0}38.5\pm 4.0\phantom{.0}$ & $\phantom{.0}30.1\pm 2.8\phantom{.0}$ \\
1300 & $>$140 & \NA & \NA &  1100--1595 & $<$15 &  $\phantom{0}0.3 \pm 0.3\phantom{0}$ & 1 & $\phantom{.0}38.5\pm 4.0\phantom{.0}$ & $\phantom{.0}30.7\pm 3.0\phantom{.0}$ \\
1400 & $>$140 & \NA & \NA &  1285--1700 & $<$15 &  $\phantom{0}0.1^{\phantom{0}+\phantom{0}0.2\phantom{0}\phantom{0}}_{\phantom{0}-\phantom{0}0.1\phantom{0}\phantom{0}}$ & 1 & $\phantom{.0}35.9\pm 3.8\phantom{.0}$ & $\phantom{.0}29.4\pm 2.8\phantom{.0}$ \\
1500 & $>$140 & \NA & \NA &  1330--1800 & $<$15 &  $\phantom{0}0.1^{\phantom{0}+\phantom{0}0.2\phantom{0}\phantom{0}}_{\phantom{0}-\phantom{0}0.1\phantom{0}\phantom{0}}$ & 1 & $\phantom{.0}36.4\pm 3.9\phantom{.0}$ & $\phantom{.0}30.0\pm 2.9\phantom{.0}$ \\
\hline
    \end{tabular}
  }
\end{table}

\begin{table}[!hptb]
  \centering
  \topcaption{
  Selection requirements on discriminating variables determined by the optimization for each Majorana neutrino mass point in the $\Pe\mu$ channel high-mass SRs. 
  Columns 7 and 8 show the total background yields (Total bkgd.) and the number of observed data (\nobs), respectively.                                                        
The last columns show the overall signal acceptance for the DY and VBF channels.
  The quoted uncertainties include both the statistical and systematic contributions.
  The dash indicates that no selection requirement is made.
  }
  \label{table:cutop_High_em}
  \cmsTable{
    \begin{tabular}{cccccccccc}
\hline
\mN & $p^{\ell_1}_{\mathrm{T}}$  & $\pt^{\ell_2}$ & $m(\ell^{\pm} \ell^{\pm} \pwjet)$  & $m(\ell \pwjet)$  & \metSQst & Total bkgd. & $\nobs$ & DY $A\epsilon$ & VBF $A\epsilon$\\
(\GeVns) & (\GeVns) & (\GeVns) & (\GeVns) & (\GeVns) & (\GeVns) & & & (\%)& (\%)\\
\hline
\textbf{\boldmath{$\Pe \mu$} channel SR1} &&&&&&&&\\
85 & $>$30 & $>$10 & $>$120 & 55--95 & $<$7 & $\phantom{}26.1 \pm 6.2\phantom{0}$ & 25 & $\phantom{0.}0.21\pm 0.03\phantom{.}$ & \NA  \\
90 & $>$30 & $>$10 & $>$120 & 60--100 & $<$7 & $\phantom{}37.4 \pm 8.4\phantom{0}$ & 32 & $\phantom{0.}0.59\pm 0.07\phantom{.}$ & \NA  \\
100 & $>$25 & $>$20 & $>$110 & 60--115 & $<$7 & $\phantom{}23.6 \pm 4.8\phantom{0}$ & 21 & $\phantom{0.0}1.3\pm 0.1\phantom{.0}$ & \NA  \\
125 & $>$30 & $>$30 & $>$140 & 90--140 & $<$7 & $\phantom{}25.5 \pm 5.9\phantom{0}$ & 16 & $\phantom{0.0}3.1\pm 0.2\phantom{.0}$ & \NA  \\
150 & $>$45 & $>$35 & $>$150 & 100--170 & $<$7 & $\phantom{}34.1 \pm 6.0\phantom{0}$ & 26 & $\phantom{0.0}5.1\pm 0.3\phantom{.0}$ & \NA  \\
200 & $>$65 & $>$35 & $>$270 & 170--230 & $<$7 & $\phantom{}11.1 \pm 2.8\phantom{0}$ & 14 & $\phantom{0.0}6.1\pm 0.4\phantom{.0}$ & \NA  \\
250 & $>$75 & $>$60 & $>$300 & 200--280 & $<$7 & $\phantom{}11.1 \pm 2.3\phantom{0}$ & 9 & $\phantom{0.0}8.9\pm 0.5\phantom{.0}$ & \NA  \\
300 & $>$95 & $>$60 & $>$340 & 255--325 & $<$7 & $\phantom{0}5.8 \pm 1.7\phantom{0}$ & 8 & $\phantom{0.0}9.0\pm 0.6\phantom{.0}$ & $\phantom{.0}3.4\pm 0.3\phantom{.0}$ \\
400 & $>$120 & $>$60 & $>$530 & 325--450 & $<$7 & $\phantom{0}2.2 \pm 1.0\phantom{0}$ & 7 & $\phantom{0.0}7.4\pm 0.4\phantom{.0}$ & $\phantom{.0}3.0\pm 0.3\phantom{.0}$ \\
500 & $>$150 & $>$60 & $>$580 & 315--530 & $<$7 & $\phantom{0}1.8 \pm 1.1\phantom{0}$ & 6 & $\phantom{0.0}6.6\pm 0.5\phantom{.0}$ & $\phantom{.0}3.0\pm 0.2\phantom{.0}$ \\
600 & $>$175 & \NA & $>$670 & 315--740 & $<$7 & $\phantom{0}1.2 \pm 0.9\phantom{0}$ & 4 & $\phantom{0.0}5.9\pm 0.4\phantom{.0}$ & $\phantom{.0}3.5\pm 0.3\phantom{.0}$ \\
700 & $>$180 & \NA & $>$720 & 350--1030 & $<$7 & $\phantom{0}1.6 \pm 1.1\phantom{0}$ & 3 & $\phantom{0.0}5.2\pm 0.3\phantom{.0}$ & $\phantom{.0}3.8\pm 0.2\phantom{.0}$ \\
800 & $>$180 & \NA & $>$720 & 400--1030 & $<$7 & $\phantom{0}1.6 \pm 1.1\phantom{0}$ & 3 & $\phantom{0.0}4.5\pm 0.3\phantom{.0}$ & $\phantom{.0}3.7\pm 0.2\phantom{.0}$ \\
900 & $>$185 & \NA & $>$720 & 450--1040 & $<$7 & $\phantom{0}1.0 \pm 0.7\phantom{0}$ & 2 & $\phantom{0.0}3.8\pm 0.2\phantom{.0}$ & $\phantom{.0}3.3\pm 0.2\phantom{.0}$ \\
1000 & $>$185 & \NA & $>$720 & 500--1415 & $<$7 & $\phantom{0}1.0 \pm 0.7\phantom{0}$ & 2 & $\phantom{0.0}3.4\pm 0.2\phantom{.0}$ & $\phantom{.0}3.0\pm 0.2\phantom{.0}$ \\
1100 & $>$185 & \NA & $>$720 & 550--1640 & $<$7 & $\phantom{0}1.0 \pm 0.7\phantom{0}$ & 1 & $\phantom{0.0}2.8\pm 0.2\phantom{.0}$ & $\phantom{.0}2.6\pm 0.2\phantom{.0}$ \\
1200 & $>$185 & \NA & $>$720 & 600--1780 & $<$7 & $\phantom{0}1.0 \pm 0.7\phantom{0}$ & 1 & $\phantom{0.0}2.4\pm 0.2\phantom{.0}$ & $\phantom{.0}2.3\pm 0.2\phantom{.0}$ \\
1300 & $>$185 & \NA & $>$720 & 650--1880 & $<$7 & $\phantom{0}0.8 \pm 0.7\phantom{0}$ & 1 & $\phantom{0.0}2.1\pm 0.1\phantom{.0}$ & $\phantom{.0}1.9\pm 0.2\phantom{.0}$ \\
1400 & $>$185 & \NA & $>$720 & 650--1885 & $<$7 & $\phantom{0}0.8 \pm 0.7\phantom{0}$ & 1 & $\phantom{0.0}1.8\pm 0.1\phantom{.0}$ & $\phantom{.0}1.7\pm 0.2\phantom{.0}$ \\
1500 & $>$185 & \NA & $>$720 & 650--1885 & $<$7 & $\phantom{0}0.8 \pm 0.7\phantom{0}$ & 1 & $\phantom{0.0}1.5\pm 0.1\phantom{.0}$ & $\phantom{.0}1.5\pm 0.1\phantom{.0}$ \\
1700 & $>$185 & \NA & $>$720 & 650--2085 & $<$7 & $\phantom{0}0.8 \pm 0.7\phantom{0}$ & 1 & $\phantom{0.0}1.2\pm 0.1\phantom{.0}$ & $\phantom{.0}1.3\pm 0.1\phantom{.0}$ \\
\textbf{\boldmath{$\Pe \mu$} channel SR2} &&&&&&\\
85 & $>$25 & $>$10 & \NA &  \NA & $<$15 &  $\phantom{}24.2 \pm 6.4\phantom{0}$ & 31 & $\phantom{0.}0.001\pm 0.002\phantom{.}$ & \NA  \\
90 & $>$25 & $>$10 & \NA &  90--240 & $<$15 &  $\phantom{}13.4 \pm 3.7\phantom{0}$ & 22 & $\phantom{0.}0.003\pm 0.002\phantom{.}$ & \NA  \\
100 & $>$30 & $>$15 & \NA &  100--335 & $<$15 &  $\phantom{}14.1 \pm 4.1\phantom{0}$ & 21 & $\phantom{0.}0.009\pm 0.003\phantom{.}$ & \NA  \\
125 & $>$35 & $>$25 & \NA &  115--150 & $<$15 &  $\phantom{0}0.6 \pm 0.4\phantom{0}$ & 2 & $\phantom{0.}0.03\pm 0.01\phantom{.}$ & \NA  \\
150 & $>$45 & $>$30 & \NA &  132--180 & $<$15 &  $\phantom{0}1.4 \pm 0.5\phantom{0}$ & 2 & $\phantom{0.}0.14\pm 0.02\phantom{.}$ & \NA  \\
200 & $>$70 & $>$30 & \NA &  180--225 & $<$15 &  $\phantom{0}1.5 \pm 0.5\phantom{0}$ & 3 & $\phantom{0.}0.86\pm 0.09\phantom{.}$ & \NA  \\
250 & $>$75 & $>$55 & \NA &  225--280 & $<$15 &  $\phantom{0}1.2 \pm 0.4\phantom{0}$ & 2 & $\phantom{0.0}1.7\pm 0.2\phantom{.0}$ & \NA  \\
300 & $>$95 & $>$55 & \NA &  280--340 & $<$15 &  $\phantom{0}1.2 \pm 0.7\phantom{0}$ & 1 & $\phantom{0.0}4.4\pm 0.4\phantom{.0}$ & $\phantom{0.0}0.8\pm 0.1\phantom{.0}$ \\
400 & $>$125 & $>$55 & \NA &  340--475 & $<$15 &  $\phantom{0}2.0 \pm 1.2\phantom{0}$ & 1 & $\phantom{.0}11.8\pm 1.1\phantom{.0}$ & $\phantom{0.0}2.7\pm 0.3\phantom{.0}$ \\
500 & $>$145 & $>$60 & \NA &  460--555 & $<$15 &  $\phantom{0}0.7 \pm 0.3\phantom{0}$ & 0 & $\phantom{.0}16.7\pm 1.6\phantom{.0}$ & $\phantom{0.0}5.2\pm 0.5\phantom{.0}$ \\
600 & $>$160 & \NA & \NA &  555--645 & $<$15 &  $\phantom{0}1.4 \pm 0.9\phantom{0}$ & 1 & $\phantom{.0}20.2\pm 1.9\phantom{.0}$ & $\phantom{.0}13.2\pm 1.2\phantom{.0}$ \\
700 & $>$170 & \NA & \NA &  610--780 & $<$15 &  $\phantom{0}2.0 \pm 0.9\phantom{0}$ & 2 & $\phantom{.0}25.0\pm 2.4\phantom{.0}$ & $\phantom{.0}17.6\pm 1.6\phantom{.0}$ \\
800 & $>$170 & \NA & \NA &  730--895 & $<$15 &  $\phantom{0}0.8 \pm 0.4\phantom{0}$ & 2 & $\phantom{.0}26.1\pm 2.5\phantom{.0}$ & $\phantom{.0}18.3\pm 1.6\phantom{.0}$ \\
900 & $>$180 & \NA & \NA &  845--1015 & $<$15 &  $\phantom{0}0.5 \pm 0.2\phantom{0}$ & 0 & $\phantom{.0}25.6\pm 2.5\phantom{.0}$ & $\phantom{.0}18.5\pm 1.7\phantom{.0}$ \\
1000 & $>$180 & \NA & \NA &  930--1075 & $<$15 &  $\phantom{0}0.2 \pm 0.2\phantom{0}$ & 0 & $\phantom{.0}23.5\pm 2.3\phantom{.0}$ & $\phantom{.0}17.6\pm 1.6\phantom{.0}$ \\
1100 & $>$180 & \NA & \NA &  1020--1340 & $<$15 &  $\phantom{0}0.3 \pm 0.3\phantom{0}$ & 0 & $\phantom{.0}26.9\pm 2.7\phantom{.0}$ & $\phantom{.0}19.6\pm 1.7\phantom{.0}$ \\
1200 & $>$180 & \NA & \NA &  1080--1340 & $<$15 &  $\phantom{0}0.1^{\phantom{0}+\phantom{0}0.2\phantom{0}\phantom{0}}_{\phantom{0}-\phantom{0}0.1\phantom{0}\phantom{0}}$ & 0 & $\phantom{.0}25.9\pm 2.6\phantom{.0}$ & $\phantom{.0}19.9\pm 1.8\phantom{.0}$ \\
1300 & $>$180 & \NA & \NA &  1155--1595 & $<$15 &  $\phantom{0}0.2^{\phantom{0}+\phantom{0}0.2\phantom{0}\phantom{0}}_{\phantom{0}-\phantom{0}0.2\phantom{0}\phantom{0}}$ & 0 & $\phantom{.0}27.1\pm 2.7\phantom{.0}$ & $\phantom{.0}20.7\pm 1.9\phantom{.0}$ \\
1400 & $>$180 & \NA & \NA &  1155--1615 & $<$15 &  $\phantom{0}0.2^{\phantom{0}+\phantom{0}0.3\phantom{0}\phantom{0}}_{\phantom{0}-\phantom{0}0.2\phantom{0}\phantom{0}}$ & 0 & $\phantom{.0}26.7\pm 2.7\phantom{.0}$ & $\phantom{.0}20.8\pm 2.0\phantom{.0}$ \\
1500 & $>$180 & \NA & \NA &  1345--1615 & $<$15 &  $\phantom{0}0.0^{\phantom{0}+\phantom{0}0.1\phantom{0}\phantom{0}}_{\phantom{0}-\phantom{0}0.0\phantom{0}\phantom{0}}$ & 0 & $\phantom{.0}21.6\pm 2.2\phantom{.0}$ & $\phantom{.0}18.0\pm 1.7\phantom{.0}$ \\
1700 & $>$180 & \NA & \NA &  1400--1800 & $<$15 &  $\phantom{0}0.7 \pm 0.6\phantom{0}$ & 0 & $\phantom{.0}19.8\pm 2.1\phantom{.0}$ & $\phantom{.0}17.0\pm 1.7\phantom{.0}$ \\
\hline
    \end{tabular}
  }
\end{table}

\section{Summary}
A search for heavy Majorana neutrinos, \N, in final states with same-sign dileptons and jets has been performed in proton-proton collisions at a center-of-mass energy of 13\TeV,
using a data set corresponding to an integrated luminosity of 35.9\fbinv.
No significant excess of events compared to the expected standard model background prediction is observed.
Upper limits at 95\% confidence level are set on the mixing matrix element between standard model neutrinos and \N ($\abs{\VlN}$) in the context of a \typeI seesaw model,
as a function of \N mass.
The analysis improves on previous 8\TeV searches by including single-jet events into the signal region,
which increases sensitivities.
For an \N mass of 40\GeV the observed (expected) limits are
$\VeNsq < 9.5\,(8.0) \times 10^{-5}$,
$\VmNsq < 2.3\,(1.9) \times 10^{-5}$,
and $\VemNsq < 2.7\,(2.7) \times 10^{-5}$,
and for an \N mass of 1000\GeV the limits are
$\VeNsq < 0.42\,(0.32)$,
$\VmNsq < 0.27\,(0.16)$,
and $\VemNsq < 0.14\,(0.14)$.
The search is sensitive to masses of \N from 20 to 1600\GeV.
The limits on the mixing matrix elements are placed up to
1240\GeV for \VeNsq,
1430\GeV for the \VmNsq, and
1600\GeV for \VemNsq.
These are the most restrictive direct limits on the \N mixing parameters for heavy Majorana neutrino masses greater than 430\GeV, and are the first for masses greater than 1200\GeV.

\begin{acknowledgments}
\hyphenation{Bundes-ministerium Forschungs-gemeinschaft Forschungs-zentren Rachada-pisek} We congratulate our colleagues in the CERN accelerator departments for the excellent performance of the LHC and thank the technical and administrative staffs at CERN and at other CMS institutes for their contributions to the success of the CMS effort. In addition, we gratefully acknowledge the computing centers and personnel of the Worldwide LHC Computing Grid for delivering so effectively the computing infrastructure essential to our analyses. Finally, we acknowledge the enduring support for the construction and operation of the LHC and the CMS detector provided by the following funding agencies: the Austrian Federal Ministry of Science, Research and Economy and the Austrian Science Fund; the Belgian Fonds de la Recherche Scientifique, and Fonds voor Wetenschappelijk Onderzoek; the Brazilian Funding Agencies (CNPq, CAPES, FAPERJ, FAPERGS, and FAPESP); the Bulgarian Ministry of Education and Science; CERN; the Chinese Academy of Sciences, Ministry of Science and Technology, and National Natural Science Foundation of China; the Colombian Funding Agency (COLCIENCIAS); the Croatian Ministry of Science, Education and Sport, and the Croatian Science Foundation; the Research Promotion Foundation, Cyprus; the Secretariat for Higher Education, Science, Technology and Innovation, Ecuador; the Ministry of Education and Research, Estonian Research Council via IUT23-4 and IUT23-6 and European Regional Development Fund, Estonia; the Academy of Finland, Finnish Ministry of Education and Culture, and Helsinki Institute of Physics; the Institut National de Physique Nucl\'eaire et de Physique des Particules~/~CNRS, and Commissariat \`a l'\'Energie Atomique et aux \'Energies Alternatives~/~CEA, France; the Bundesministerium f\"ur Bildung und Forschung, Deutsche Forschungsgemeinschaft, and Helmholtz-Gemeinschaft Deutscher Forschungszentren, Germany; the General Secretariat for Research and Technology, Greece; the National Research, Development and Innovation Fund, Hungary; the Department of Atomic Energy and the Department of Science and Technology, India; the Institute for Studies in Theoretical Physics and Mathematics, Iran; the Science Foundation, Ireland; the Istituto Nazionale di Fisica Nucleare, Italy; the Ministry of Science, ICT and Future Planning, and National Research Foundation (NRF), Republic of Korea; the Lithuanian Academy of Sciences; the Ministry of Education, and University of Malaya (Malaysia); the Ministry of Science of Montenegro; the Mexican Funding Agencies (BUAP, CINVESTAV, CONACYT, LNS, SEP, and UASLP-FAI); the Ministry of Business, Innovation and Employment, New Zealand; the Pakistan Atomic Energy Commission; the Ministry of Science and Higher Education and the National Science Center, Poland; the Funda\c{c}\~ao para a Ci\^encia e a Tecnologia, Portugal; JINR, Dubna; the Ministry of Education and Science of the Russian Federation, the Federal Agency of Atomic Energy of the Russian Federation, Russian Academy of Sciences, the Russian Foundation for Basic Research, and the National Research Center ``Kurchatov Institute"; the Ministry of Education, Science and Technological Development of Serbia; the Secretar\'{\i}a de Estado de Investigaci\'on, Desarrollo e Innovaci\'on, Programa Consolider-Ingenio 2010, Plan Estatal de Investigaci\'on Cient\'{\i}fica y T\'ecnica y de Innovaci\'on 2013-2016, Plan de Ciencia, Tecnolog\'{i}a e Innovaci\'on 2013-2017 del Principado de Asturias, and Fondo Europeo de Desarrollo Regional, Spain; the Ministry of Science, Technology and Research, Sri Lanka; the Swiss Funding Agencies (ETH Board, ETH Zurich, PSI, SNF, UniZH, Canton Zurich, and SER); the Ministry of Science and Technology, Taipei; the Thailand Center of Excellence in Physics, the Institute for the Promotion of Teaching Science and Technology of Thailand, Special Task Force for Activating Research and the National Science and Technology Development Agency of Thailand; the Scientific and Technical Research Council of Turkey, and Turkish Atomic Energy Authority; the National Academy of Sciences of Ukraine, and State Fund for Fundamental Researches, Ukraine; the Science and Technology Facilities Council, UK; the US Department of Energy, and the US National Science Foundation.

Individuals have received support from the Marie-Curie program and the European Research Council and Horizon 2020 Grant, contract No. 675440 (European Union); the Leventis Foundation; the A. P. Sloan Foundation; the Alexander von Humboldt Foundation; the Belgian Federal Science Policy Office; the Fonds pour la Formation \`a la Recherche dans l'Industrie et dans l'Agriculture (FRIA-Belgium); the Agentschap voor Innovatie door Wetenschap en Technologie (IWT-Belgium); the F.R.S.-FNRS and FWO (Belgium) under the ``Excellence of Science - EOS" - be.h project n. 30820817; the Ministry of Education, Youth and Sports (MEYS) of the Czech Republic; the Lend\"ulet (``Momentum") Program and the J\'anos Bolyai Research Scholarship of the Hungarian Academy of Sciences, the New National Excellence Program \'UNKP, the NKFIA research grants 123842, 123959, 124845, 124850 and 125105 (Hungary); the Council of Scientific and Industrial Research, India; the HOMING PLUS program of the Foundation for Polish Science, cofinanced from European Union, Regional Development Fund, the Mobility Plus program of the Ministry of Science and Higher Education, the National Science Center (Poland), contracts Harmonia 2014/14/M/ST2/00428, Opus 2014/13/B/ST2/02543, 2014/15/B/ST2/03998, and 2015/19/B/ST2/02861, Sonata-bis 2012/07/E/ST2/01406; the National Priorities Research Program by Qatar National Research Fund; the Programa de Excelencia Mar\'{i}a de Maeztu, and the Programa Severo Ochoa del Principado de Asturias; the Thalis and Aristeia programs cofinanced by EU-ESF, and the Greek NSRF; the Rachadapisek Sompot Fund for Postdoctoral Fellowship, Chulalongkorn University, and the Chulalongkorn Academic into Its 2nd Century Project Advancement Project (Thailand); the Welch Foundation, contract C-1845; and the Weston Havens Foundation (USA).
\end{acknowledgments}

\bibliography{auto_generated}

\cleardoublepage \appendix\section{The CMS Collaboration \label{app:collab}}\begin{sloppypar}\hyphenpenalty=5000\widowpenalty=500\clubpenalty=5000\input{EXO-17-028-authorlist.tex}\end{sloppypar}
\end{document}

%% file: EXO-17-028-authorlist.tex
\vskip\cmsinstskip
\textbf{Yerevan Physics Institute, Yerevan, Armenia}\\*[0pt]
A.M.~Sirunyan, A.~Tumasyan
\vskip\cmsinstskip
\textbf{Institut f\"{u}r Hochenergiephysik, Wien, Austria}\\*[0pt]
W.~Adam, F.~Ambrogi, E.~Asilar, T.~Bergauer, J.~Brandstetter, M.~Dragicevic, J.~Er\"{o}, A.~Escalante~Del~Valle, M.~Flechl, R.~Fr\"{u}hwirth\cmsAuthorMark{1}, V.M.~Ghete, J.~Hrubec, M.~Jeitler\cmsAuthorMark{1}, N.~Krammer, I.~Kr\"{a}tschmer, D.~Liko, T.~Madlener, I.~Mikulec, N.~Rad, H.~Rohringer, J.~Schieck\cmsAuthorMark{1}, R.~Sch\"{o}fbeck, M.~Spanring, D.~Spitzbart, A.~Taurok, W.~Waltenberger, J.~Wittmann, C.-E.~Wulz\cmsAuthorMark{1}, M.~Zarucki
% ...
\vskip\cmsinstskip
% ...
1:  Also at Vienna University of Technology, Vienna, Austria\\
% ...
